\documentclass[10pt,twocolumn,letterpaper]{article}

\usepackage{iccv}

\usepackage{listings}

\usepackage{times}
\usepackage{multicol}

\usepackage{caption} 
\usepackage{epsfig}
\usepackage{graphicx}
\usepackage{amsmath}
\usepackage{mathrsfs}
\usepackage{amssymb}
\usepackage{subfigure}
\usepackage{enumitem}

\usepackage{textcomp}
\usepackage{boldline}

\usepackage{booktabs}
\usepackage[normalem]{ulem}
\useunder{\uline}{\ul}{}
\usepackage{multirow}

\usepackage[dvipsnames]{xcolor}

\usepackage{stmaryrd}
\usepackage{trimclip}
\makeatletter
\DeclareRobustCommand{\rarr}{%
  \mathrel{\mathpalette\short@to\relax}%
}

\newcommand{\short@to}[2]{%
  \mkern3mu
  \clipbox{{.3\width} 0 0 0}{$\m@th#1\vphantom{+}{\shortrightarrow}$}%
  }
\makeatother

\usepackage{algorithm}
\usepackage{algorithmic}

\usepackage{xcolor}

\usepackage{macros}

\usepackage{xr-hyper}

\makeatletter
\newcommand*{\addFileDependency}[1]{
  \typeout{(#1)}
  \@addtofilelist{#1}
  \IfFileExists{#1}{}{\typeout{No file #1.}}
}
\makeatother

\newcommand*{\myexternaldocument}[1]{
    \externaldocument{#1}
    \addFileDependency{#1.tex}
    \addFileDependency{#1.aux}
}

\myexternaldocument{appendix}

\usepackage[pagebackref=true,breaklinks=true,letterpaper=true,colorlinks,bookmarks=false]{hyperref}
\iccvfinalcopy

\usepackage{footmisc}
\begin{document}

\title{ELF-VC: Efficient Learned Flexible-Rate Video Coding}
\author{Oren Rippel\thanks{Equal contribution},\, Alexander G. Anderson\footnotemark[1],\, Kedar Tatwawadi, Sanjay Nair, Craig Lytle, Lubomir Bourdev\\    
WaveOne, Inc.\\
{\tt\small \{oren, alex, kedar, sanjay, craig, lubomir\}@wave.one}}
\maketitle

\thispagestyle{empty}

\begin{abstract}
While learned video codecs have demonstrated great promise, they have yet to achieve sufficient efficiency for practical deployment. In this work, we propose several novel ideas for learned video compression which allow for improved performance for the low-latency mode (I- and P-frames only) along with a considerable increase in computational efficiency. In this setting, for natural videos our approach compares favorably across the entire R-D curve under metrics PSNR, MS-SSIM and VMAF against all mainstream video standards (H.264, H.265, AV1) and all ML codecs. At the same time, our approach runs at least 5x faster and has fewer parameters than all ML codecs which report these figures.

Our contributions include a flexible-rate framework allowing a single model to cover a large and dense range of bitrates, at a negligible increase in computation and parameter count;  an efficient backbone optimized for ML-based codecs; and a novel in-loop flow prediction scheme which leverages prior information towards more efficient compression.

We benchmark our method, which we call ELF-VC (Efficient, Learned and Flexible Video Coding) on popular video test sets UVG and MCL-JCV under metrics PSNR, MS-SSIM and VMAF. For example, on UVG under PSNR, it reduces the BD-rate by 44\% against H.264, 26\% against H.265, 15\% against AV1, and 35\% against the current best ML codec. At the same time, on an NVIDIA Titan V GPU our approach encodes/decodes VGA at 49/91 FPS, HD 720 at 19/35 FPS, and HD 1080 at 10/18 FPS.
\end{abstract}

\section{Introduction}

The trends of growth of video capture and consumption are staggering. Every day, 1.5 billion hours of videos are watched across YouTube, Netflix and Facebook, and 23 million new cameras are added into circulation \cite{youtube,netflix,facebook,ldv2017}. 

In the last few years, ML-based compression algorithms have shown promise in their potential to mitigate some of this global video congestion. The ML subfield of end-to-end methods for image compression has grown rapidly with hundreds of papers \cite[\dots]{balle2016end,rippel17,balle2018variational} which demonstrate unequivocally that learned approaches can achieve improved coding efficiency relative to their hard-coded counterparts. 

\begin{figure}[t]
    \centering
    \includegraphics[height=1.64in]{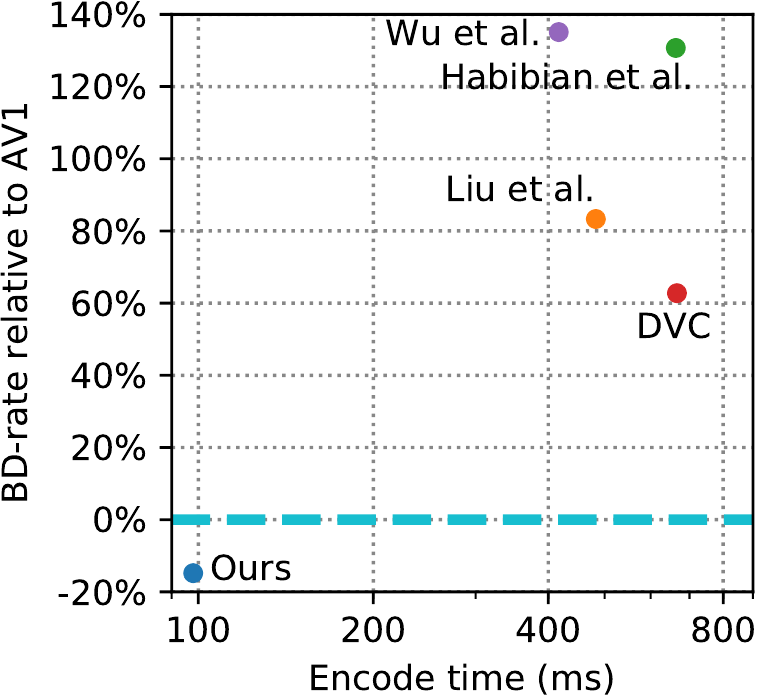}\hspace{-0.16in}
    \includegraphics[height=1.64in]{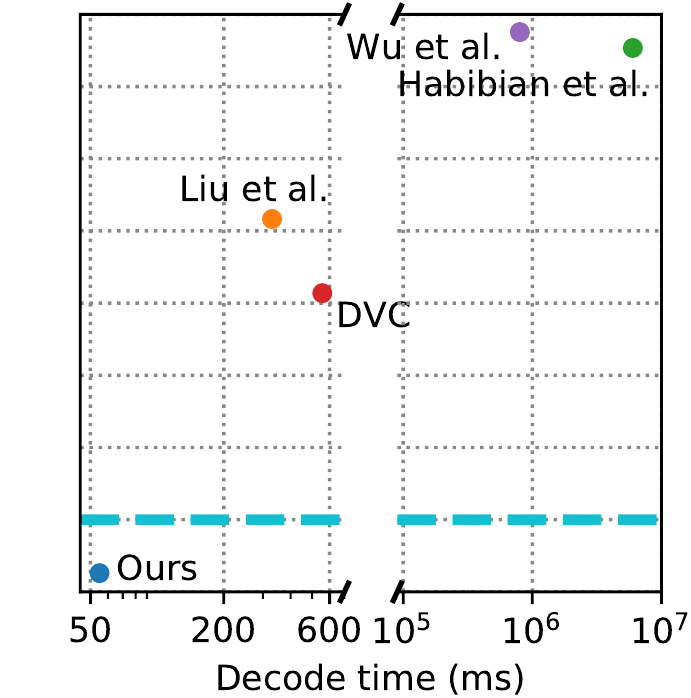}
    \caption{BD-Rate for ML-based codecs relative to AV1 as a function of encode/decode time on HD 1080 videos \cite{wu2018vcii,habibian2019video,lu2019dvc,liu2020fast} (UVG dataset, PSNR metric). Our approach reduces the BD-rate by 54\% relative to the current fastest ML codec which reports speed \cite{liu2020fast}, while running 5x faster.}
    \label{fig:intro_eyecandy}
\end{figure}

These approaches have, in turn, planted the seeds for ML-based video coding algorithms. Even though end-to-end video coding research has only taken its first few steps, it is clear that learned approaches have the potential to yield significant bitrate savings over the existing standards \cite[\dots]{rippel2018learned,lu2019dvc,habibian2019video,golinski2020feedback,lu2020content}. However, there still exists an elephant in the room: is it possible for ML-based approaches to achieve sufficient flexibility and efficiency to become practical in the real world? 

We propose a new ML video codec, ELF-VC (Efficient, Learned and Flexible-Rate Video Coding) for the low-latency mode, which aims to improve three key weaknesses of ML-based video compression: bitrate flexibility, compression efficiency, and speed.

\paragraph{Bitrate flexibility} Traditional codecs can dynamically adjust the bitrate to achieve a target bandwidth or target compression quality as a function of the complexity of the video and changing network conditions. Most existing ML codecs, however, represent each point on the R-D curve with a separate model, which is impractical due to parameter explosion and model loading inefficiency. In contrast, ELF is able to support a large and dense range of bitrates on a per-frame basis with a single set of parameters. 

\paragraph{Compression Efficiency} While ML-based codecs have shown improved compression efficiency over H.265, no ML codec has yet to outperform the standards across the entire PSNR curve. Moreover, no benchmarks have been presented on the VMAF metric, nor against AV1, as ML codecs have not compared favorably in these settings. We benchmark ELF on popular video test sets UVG and MCL-JCV under metrics PSNR, MS-SSIM and VMAF. In the low-latency mode, for natural videos ELF compares favorably across the entire R-D curve against the standards, and all other ML codecs. For example, on UVG under PSNR, ELF reduces the BD-rate by $44\%$ against H.264, $26\%$ against H.265, $15\%$ against AV1, $35\%$ against the next-best ML codec \cite{agustsson2020scale}, and 54\% against the next-fastest ML-codec \cite{liu2020fast} (while running 5x faster).

\paragraph{Speed} For any practical application encoding must run at a reasonable frame rate, and decoding must run in real-time. Research on learned video compression has focused on improving the R-D curve, often at the expense of speed. For instance, autoregressive approaches inherently cannot be parallelized, resulting in methods that take many seconds to decode a single frame. In contrast, ELF runs at least 5x faster than all other ML codecs which report timings, and with fewer parameters. On an NVIDIA Titan V GPU, ELF encodes/decodes VGA at 49/91 FPS, HD 720 at 19/35 FPS, and HD 1080 at 10/18 FPS.

\paragraph{}Our primary contributions are:
\begin{enumerate}
\item A novel framework for efficient rate control for learned video coding. This allows a single model to encode each frame with a wide range of bitrates, at a negligible increase to computation and parameter count. 
\item A backbone specifically optimized to achieve strong performance on compression tasks while remaining computationally efficient.
\item An \emph{in-loop flow predictor}, a novel module that utilizes previously transmitted information to get a strong initial estimate of the motion for the current frame.
\end{enumerate}

\begin{figure*}[t]
    \centering
    \includegraphics[width=\textwidth]{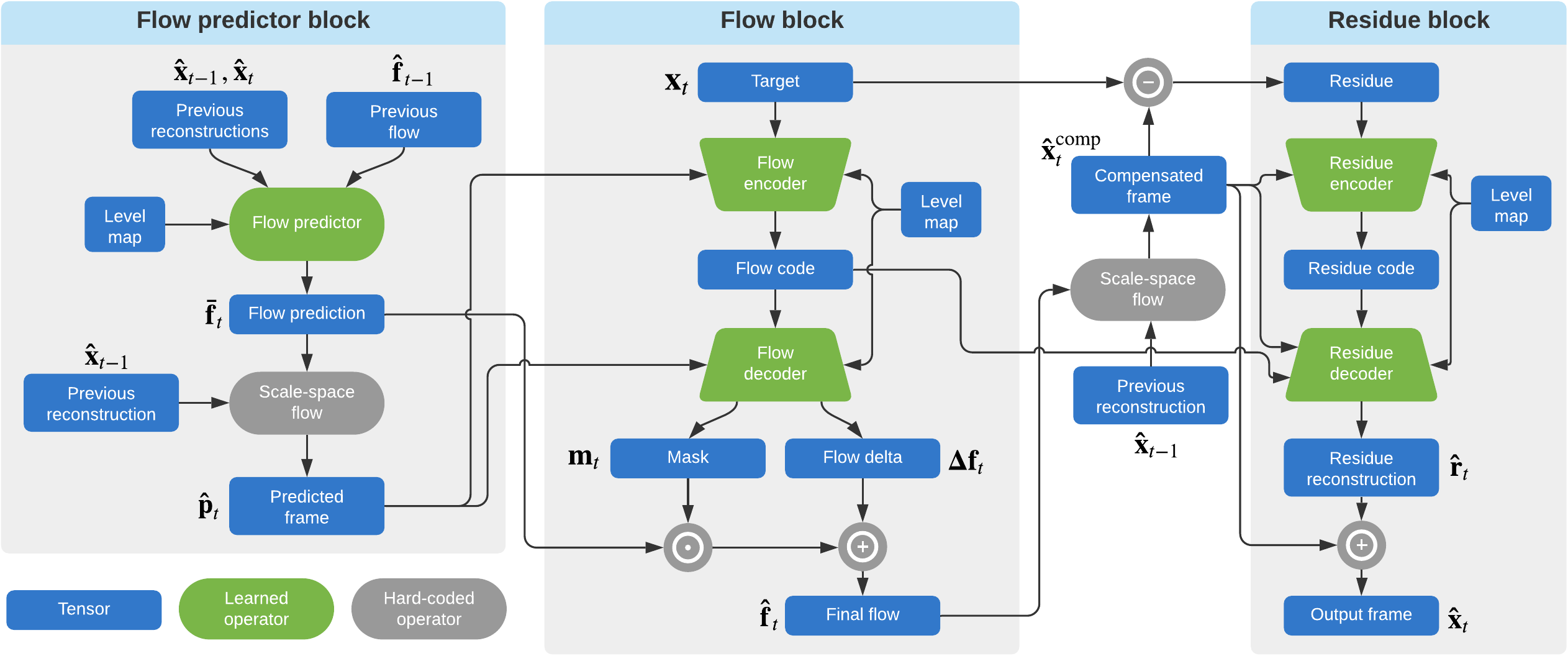}
    \caption{The overall architecture of ELF-VC. The {\bf predictor block} (Section \ref{sec:predictor}) uses previously transmitted information to get a strong initial estimate of the motion for the current frame, without sending any bits. The {\bf flow block} refines this motion estimate and the {\bf residue block} reconstructs the remaining residual. The {\bf level map} is provided as input to different parts of the model to facilitate conditional modeling (Section \ref{sec:flexible_rate_framework}). The learned operators are powered by the neural network backbone  described in Section \ref{sec:backbone}. The details of the state propagation and layer specifications are omitted for clarity and can be found Appendix \ref{sec:backbone_details}.}
    \label{fig:overall_architecture}
\end{figure*}

\subsection{Related work}\label{sec:related_work}
\paragraph{ML-based flexible-rate compression}
Compared to previous work, this paper is the first efficient flexible-rate video compression paper. Beyond the loss formulation, the fundamental challenge of learned flexible-rate compression is supporting a large dynamic range of bitrates in a single model. As the range gets larger and denser, performance often degrades. The topic of flexible-rate modeling has been addressed in a number of previous learned image coding papers which use conditional convolution \cite{Choi2019}, variable quantization width \cite{Cui2020, Guo2020}, and recurrent networks \cite{toderici2015variable}. While the former methods are slow due to an autoregressive probability model for the codelayer or a recurrent network, our work doesn't use such methods. 

In the learned video compression literature, \cite{rippel2018learned} use a spatial multiplexer to achieve flexible rates with a single model. However, this approach requires a slow and complex search at encode-time in order to produce the spatial multiplexer map.

\paragraph{Traditional video compression}
There has been a long history of hand-designed video codecs, such as H.263 and H.264 \cite{wiegand2003overview}, which form the basis of video standards widely used today. More recently, H.265 \cite{sullivan2012overview}, VP9 \cite{mukherjee2013latest} and AV1 \cite{chen2018overview} have made significant improvements over the legacy video standards, and continue to be an active research area. The traditional codecs have been exceptionally well-engineered and tuned, and have been difficult for the ML community to match both in terms of compression fidelity but also computational efficiency.

\paragraph{ML-based video compression}
Due to the inherent complexity of designing video compression algorithms, various works formulate end-to-end solutions to conquer subsets of the grand problem. One class of approaches such as \cite{wu2018vcii,djelouah2019neural,park2019deep} focuses directly on interpolation and omit P-frames. Another popular direction (which this work follows) is to design a low-latency ML-based codec, which only features keyframe compression and forward frame extrapolation (i.e I/P-frames only) \cite{lu2019dvc,liu2019neural,liu2020fast}. Promising recent directions involve modeling motion using scale-space flow \cite{agustsson2020scale} and resolution-adaptive flow \cite{liu2020nvc,hu2020ECCV}, propagating a latent state \cite{rippel2018learned,golinski2020feedback}, and explicitly mitigating error propagation \cite{lu2020content}. Yet another promising approach \cite{habibian2019video} revolves around using spatiotemporal autoencoders to encode chunks of frames.

\paragraph{Efficient ML compression}
Research on computationally efficient ML codecs is still in its nascence. In the image compression world, several approaches explore efficient codec modeling, relying on architecture optimizations \cite{rippel17,johnston2019computationally,liu2020unified}. To our knowledge, the only other ML video codec work with focus on the topic is \cite{liu2020fast}, whose approach is to remove inter-dependency between frames and instead rely on entropy conditioning to capture redundancy.

\section{Problem setup and baseline model}
\label{sec:baseline_model}
We aim to encode a video with frames $\rmbx_1, \ldots, \rmbx_T\in[0, 1]^{3\times H\times W}$ using the low-latency mode including I- and P-frames only. In this section, we describe a baseline model which is a combination of ideas from previous state-of-the-art ML codecs.

\paragraph{Baseline I-frame model} 
The keyframe compressor (I-frame) is an image codec which encodes individual frames. As discussed in Section \ref{sec:related_work}, there is a well-established body of work on learned techniques for image compression. For the baseline I-frame model, we use the backbone presented in Section \ref{sec:backbone}, combined with a simplified variant of the hyperprior coding scheme presented in \cite{balle2018variational} (see Appendix \ref{sec:codelayer_details} for details) and avoid any context modeling.

\paragraph{Baseline P-frame model} 
The starting point for our baseline P-frame model is the common flow-residue model \cite[\dots]{lu2019dvc,agustsson2018generative}, where an autoencoder (``flow block'') is used to reconstruct an optical flow field, and the \emph{residue} --- the leftover difference with the target --- is compressed with a second autoencoder (``residue block''). Similar to \cite{rippel2018learned}, we propagate a state across blocks and across frames, which results in a significant improvement in the model (Table \ref{tab:ablation}). 

The flow block is an autoencoder which takes in the target $\rmbx_t$, previous reconstruction $\rmbhx_{t-1}$ and previous state $\rmbs_{t-1}$ as inputs. Through a bottleneck, it updates the state and then produces a flow output $\rmbhf_t$ to compensate from $\rmbhx_{t-1}$ to $\rmbx_t$ using the warping transform $\rmbF(\rmbhx_{t-1}, \rmbhf_t)$. We build on \cite{agustsson2020scale} and for $\rmbF(\cdot, \cdot)$ adopt the scale-space flow operator which elegantly handles uncertainty in motion estimation. Such a flow $\rmbhf_t$ is made of 3 channels: horizontal and vertical displacements along with a blurring parameter (scale-space sigma). The compensated frame is then $\rmbhx^{\textrm{comp}}_t=\rmbF(\rmbhx_{t-1}, \rmbhf_t)$.

The residue block then takes in the leftover signal $\rmbx_t - \rmbhx^{\textrm{comp}}_t$ and the state returned by the flow block to produce a residue $\rmbhr_t$ which is added to the warping to produce final reconstruction $\rmbhx_t = \rmbhx^{\textrm{comp}}_t + \rmbhr_t$. It also outputs a final state $\rmbs_t$ that is passed to the next frame.

The architecture for this baseline is essentially the flow and residual blocks in Figure \ref{fig:overall_architecture} without level maps. The backbones for both blocks are presented in \ref{sec:backbone}. Each block of the P-frame model uses the same hyperprior coding scheme as the I-frame model. Similarly to the I-frame model, autoregressive context models are avoided because they are are prohibitively slow for practical use in their current formulation.

\section{Novel Contributions} \label{sec:approach}
In this section, we describe new ideas building on top of the baseline model (Section \ref{sec:baseline_model}). Section \ref{sec:ablation_studies} presents ablation studies of their individual contributions.

\subsection{Flexible-rate framework for ML video codecs}\label{sec:flexible_rate_framework}
Rate control, or the optimization of visual quality under real-world constraints (e.g. bandwidth and latency), is essential for the practical deployment of a video codec. For example, in order to minimize bandwidth-induced latency, it is important to constrain the average bitrate of the video \textit{and} the maximum bitrate for any frame.

While there exist methods for flexible-rate ML-based image compression (Sec. \ref{sec:related_work}), it is challenging to extend these to video without suffering a loss of performance. Our work introduces two main innovations: a novel loss modulation scheme which improves BD-rate by 10\% (see \ref{sec:ablation_studies}) by better training the model for high rates, and a novel embedding scheme which allows the rate to be smoothly varied across the rate-distortion curve. 

For each frame we initially aim to support $L$ different points on the R-D curve (we refer to them as levels) using a single model. In order for the network to optimize a level-dependent loss, the discrete level is converted into a one-hot vector of dimension $L$. This vector is tiled spatially and concatenated to the input of each of the neural network encoders and decoders ({\bf level map}, Figure \ref{fig:overall_architecture}), and as an input to the in-loop flow predictor (Sec. \ref{sec:predictor}). The I-frame model also uses a learned level and channel-dependent quantization width for the quantized codelayer, similar to \cite{Cui2020}. Variable quantization width was not sufficient to achieve competitive performance --- we found that giving the rate as input to all of the listed places to the network was necessary for optimal performance. During inference time, without any additional training, we achieve a denser sampling of the bitrate range at encode-time by linearly interpolating the $L$-dimensional level vector (see Appendix \ref{sec:rate_interpolation}). 

\begin{figure}[b]
    \centering
    \includegraphics[height=1.06in]{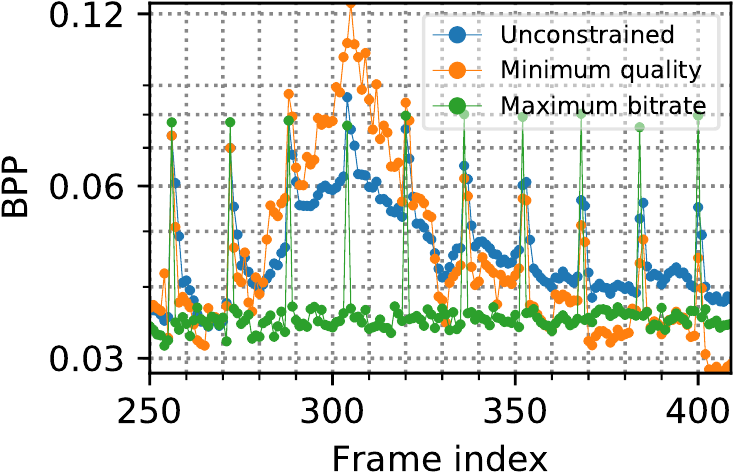}\hspace{0.02in}
    \includegraphics[height=1.06in]{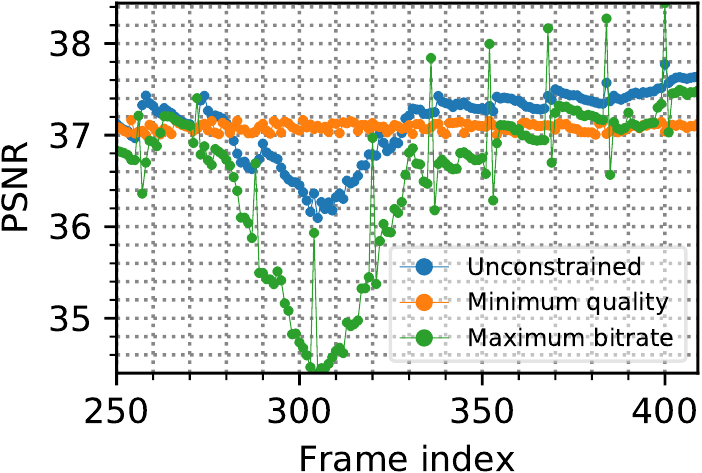}
    \caption{The flexible-rate nature of ELF facilitates deployment in the presence of different constraints. We present examples of 3 different rate controllers: maximum BPP (caps of 0.081 on I, 0.037 on P), minimum quality (PSNR of 37), and constant level. A complex event around frame 300 (Jockey video, UVG dataset) causes the BPP to spike under guaranteed minimum quality (orange curve). If we constrain the bandwidth, the quality drops (green curve).}
    \label{fig:rate_control}
\end{figure}

\paragraph{Multi-level loss setup}
The multi-level loss consists of the typical rate and distortion terms, with added level conditioning:
\begin{align}
    \mathcal{L}_{\textrm{rec}} & = \frac{1}{T} \sum_{t} \bbE_l \left[\ui{\mu_t}{l} D(\rmbx_t, \ui{\rmbhx_t}{l})\right] \\
    \mathcal{L}_{\textrm{ent}} & = \frac{1}{T} \sum_{t} \bbE_l {\ui{\lambda_\textrm{reg}}{l}} \ui{R_t}{l}\; .
\end{align}
The compression level $l$ for frame $t$ is sampled during training (see \ref{sec:experimental_setup}). $D(\cdot, \cdot)$ is the distortion metric (e.g. MSE loss), and $\ui{R_t}{l}$ is the total codelength of the encoded frame (see Appendix \ref{sec:codelayer_details}) for the flow and residual blocks. $\ui{\mu_t}{l}$ is a dynamically-chosen weight that substantially improves multi-level training as explained below (see Table \ref{tab:ablation}). The loss encourages the model to achieve different points on the R-D curve by using different regularization weights $\ui{\lambda_{\textrm{reg}}}{l}$ for different levels. 

\paragraph{Dynamic loss modulation} \label{sec:loss_modulation}
During training, the reconstruction loss for each frame and level is multiplied by a dynamically-changing weight $\ui{\mu_t}{l} \ge 1$ in order to encourage better performance for the levels and frames that are under-performing. The motivation for this idea is that when training the baseline unmodulated models (with $\ui{\mu_t}{l} = 1$), we observed that the higher bitrate levels of the variable-rate model trained much more slowly than lower ones. We further noticed that frames later in the GOP train more slowly than the earlier frames. 

\begin{figure}[t]
    \centering
    \includegraphics[width=\columnwidth]{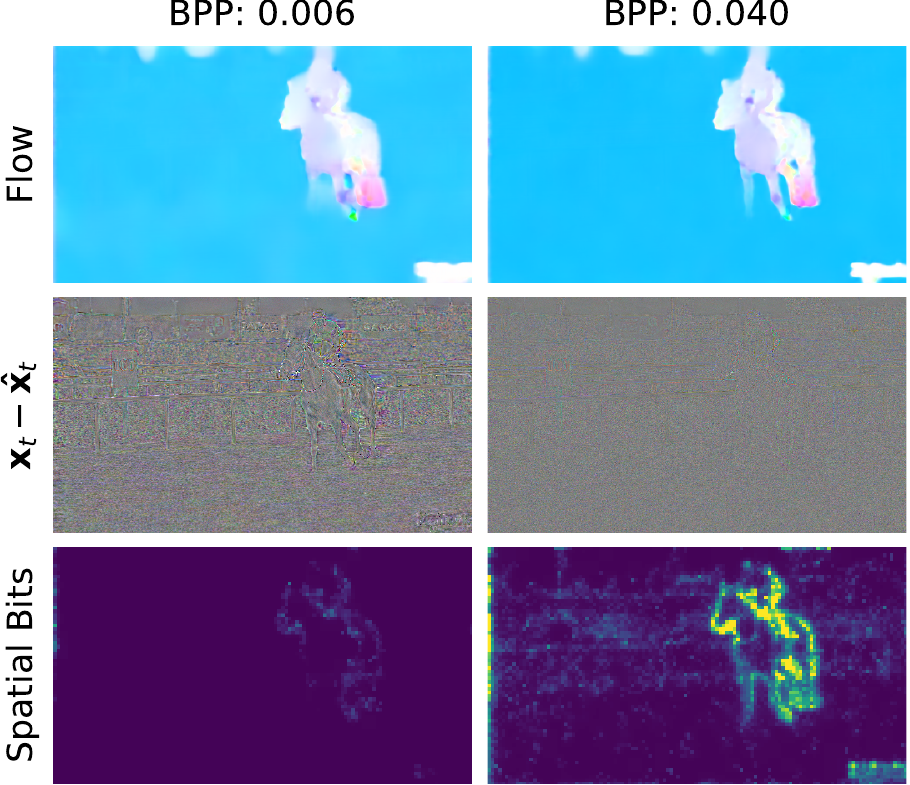}
    \caption{Visual comparison for an HD 1080 video encoded using the same model at two different levels. 
    The {\bf top row} shows the final flow $\rmbhf_t$, the {\bf middle row} shows difference between the target and the final reconstruction, and the {\bf bottom row} shows the spatial bit allocation (Appendix \ref{sec:codelayer_details}). The error is boosted by a factor of 4 and clipped for visualization purposes. The spatial bits plots use the same color mapping where yellow corresponds more bits spent.}
    \label{fig:rate_visual_comparison}
\end{figure}

\begin{figure*}[t]
    \centering
    \includegraphics[width=\textwidth]{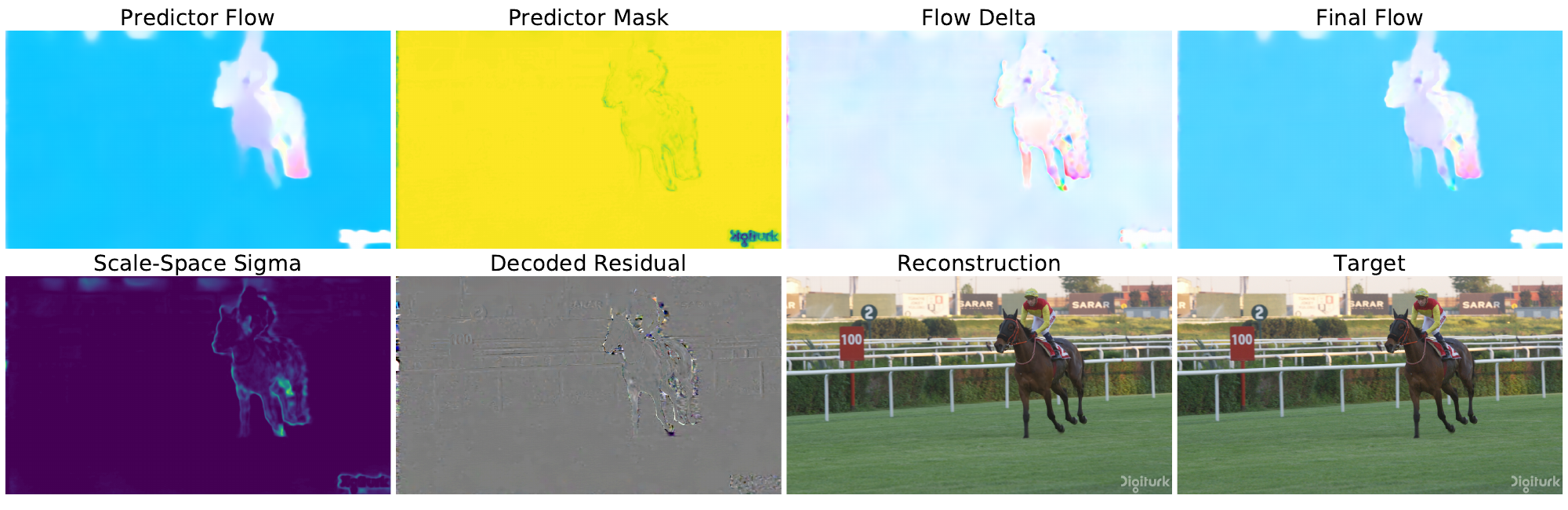}
    \caption{Intermediate tensors in the P-frame decoder for a typical HD 1080 video. The in-loop flow predictor uses previously transmitted information to generate an initial estimate of the motion ({\bf predictor flow}), which is then masked by the {\bf predictor mask}. A {\bf flow delta} is decoded from the state in the first block and added to the predicted flow to generate the {\bf final flow}. It can be seen that the predicted flow, which was computed without transmitting any additional bits, captures much of the motion --- allowing for a sparser flow delta. The flow and the {\bf scale-space sigma} (also decoded from the state) are applied to the previous frame to generate a compensated frame. The last block generates an estimate of the remaining {\bf residue} and adds it to the compensated frame to generate the {\bf final reconstruction}. The residue is multiplied by a factor of 4 and clipped for visualization purposes. All flows are normalized so that a fully saturated color is a flow with a magnitude of 15 pixels.}
    \label{fig:intermediate_tensors}
    \vspace{-0.1in}
\end{figure*}

One natural idea for the loss modulation was to increase the weight associated with the parts of the model that were under-performing. We developed the following method in order to avoid needing to set the weights manually. We fix the I-frame weights $\ui{\mu_0}{l}$ for each level to be 1. Then for a particular level, we increase the P-frame weight if that P-frame is under-performing relative to the I-frame for that level. Specifically, the moving averages $\bbE \ui{D_t}{l}$ for the $t$-th frame and $l$-th level are computed. If $\ui{\textrm{PSNR}_t}{l} < \ui{\textrm{PSNR}_0}{l} - \delta$, then $\ui{\mu_t}{l}$ is increased by a small value. If the opposite is true, $\ui{\mu_t}{l}$ is decreased. See the ablation studies \ref{sec:ablation_studies} and Appendix \ref{sec:extended_loss_modulator} for additional details and discussion.

\paragraph{Level embedding}
While a discrete set of $L$ levels is useful for accumulating moving averages to modulate the loss during training, for practical purposes typical values of $L\approx 8$ are too coarse for precise rate control. We introduce a technique which allows smoothly varying the bitrate at encode time. Given the necessity of feeding the rate to multiple parts of the model, in addition to varying the quantization width we further need to continuously interpolate the one-hot rate. To that end, linear interpolation in the embedding space was used to target intermediate rates. Previous work in the image compression world \cite{Choi2019} that used two knobs (discrete rate and quantization width) resulted in non-monotonic behavior. In contrast, our method has one continuous knob that results in the quality increasing monotonically as the bitrate increases. Furthermore, our method does not require any additional training in order to interpolate the rate with competitive performance. The detailed equations can be found in Appendix \ref{sec:rate_interpolation}.

\begin{figure}[b]
    \centering
    \includegraphics[width=0.7\columnwidth]{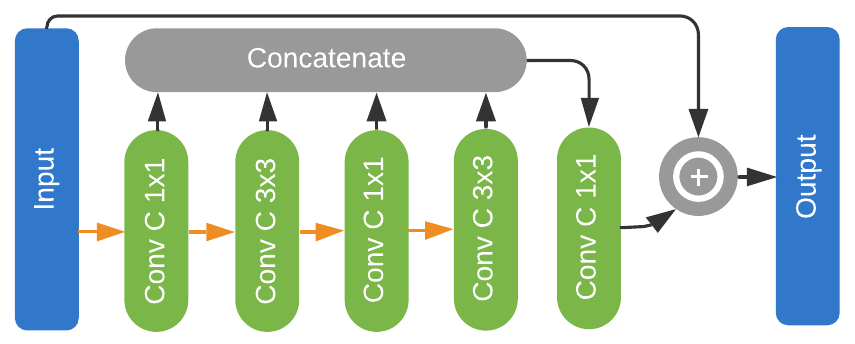}
    \caption{The Delayed Merge (DM-$C$) block with $C$ base channels. Our CNN backbones use this as the primary building block with processing across different scales. The DM block allows for the benefits of DenseNet while being faster due to the reduced number of concatenation operations (Section \ref{sec:backbone}).}
    \label{fig:delayed_merge}
\end{figure}

\subsection{Compression-centric backbone}
\label{sec:backbone}
We propose an efficient backbone that achieves competitive results for ML-based coding, which lowers the BD-rate by 30\% and improves speed by 75\% (Table \ref{sec:ablation_studies}) relative to popular video backbones. This is based on a block that we refer to as the Delayed Merge (DM) block (Figure \ref{fig:delayed_merge}). We experimented with various common backbones such as DenseNet \cite{huang2017densely}, ResNet \cite{he2016deep}, multiscale dual path \cite{chen2017dual,huang2018multiscale,rippel2018learned}, and inverted residual modules \cite{sandler2018mobilenetv2}, and found that a DenseNet-like (and optionally multiscale) representation achieves a balanced tradeoff between expressivity and computational performance. However, in its original formulation, DenseNet features many concatenation operations and convolutions with a small number of filters. As such, it does not lend itself to efficient computation. This issue also affects multiscale approaches presented in \cite{huang2018multiscale,rippel2018learned}. 

\begin{figure*}[t]
    \rotatebox[origin=t]{90}{\bf UVG Dataset\hspace{-1.5in}}\hspace{0.02in}
    \includegraphics[height=1.72in]{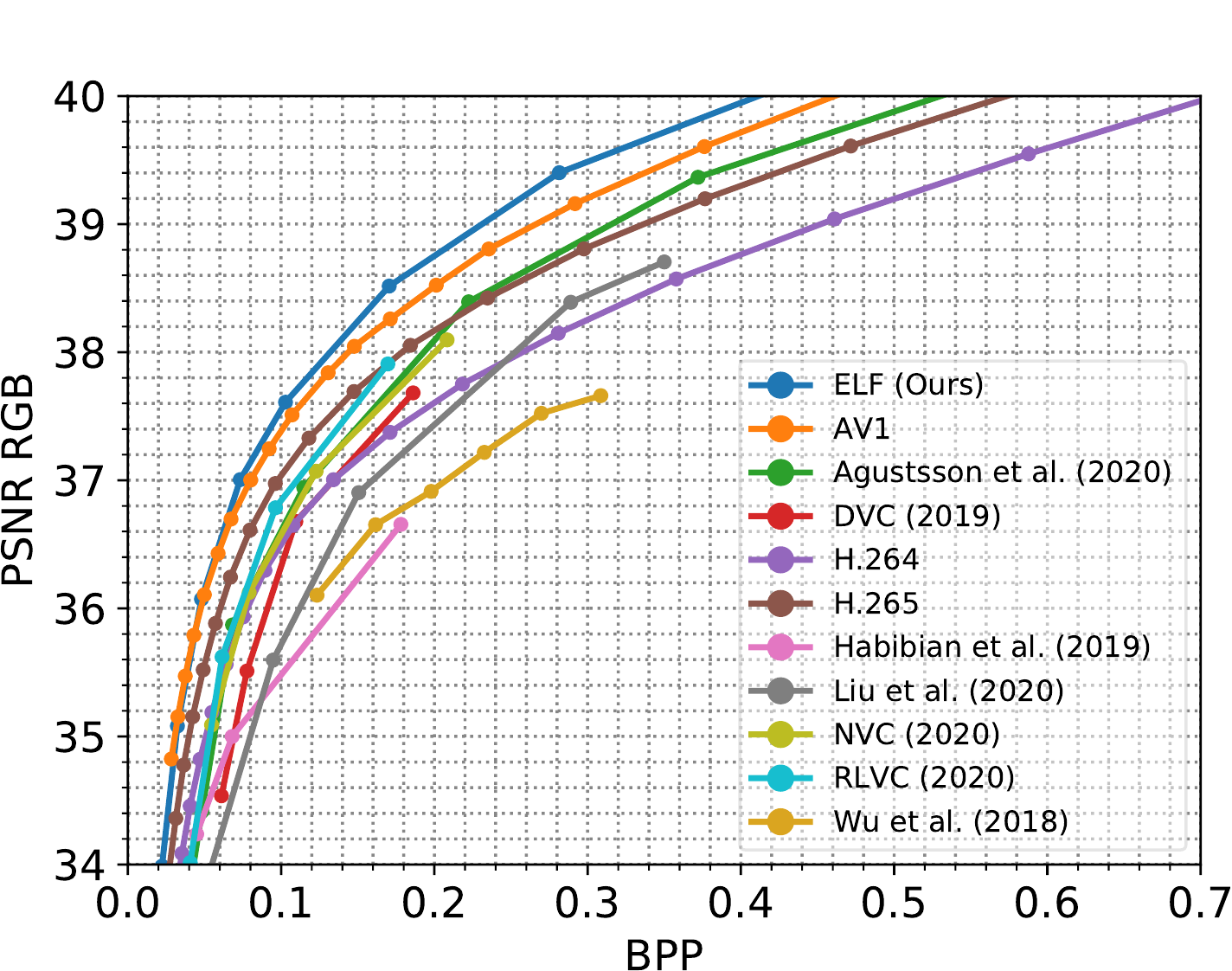}\hspace{-0.02in}
    \includegraphics[height=1.72in]{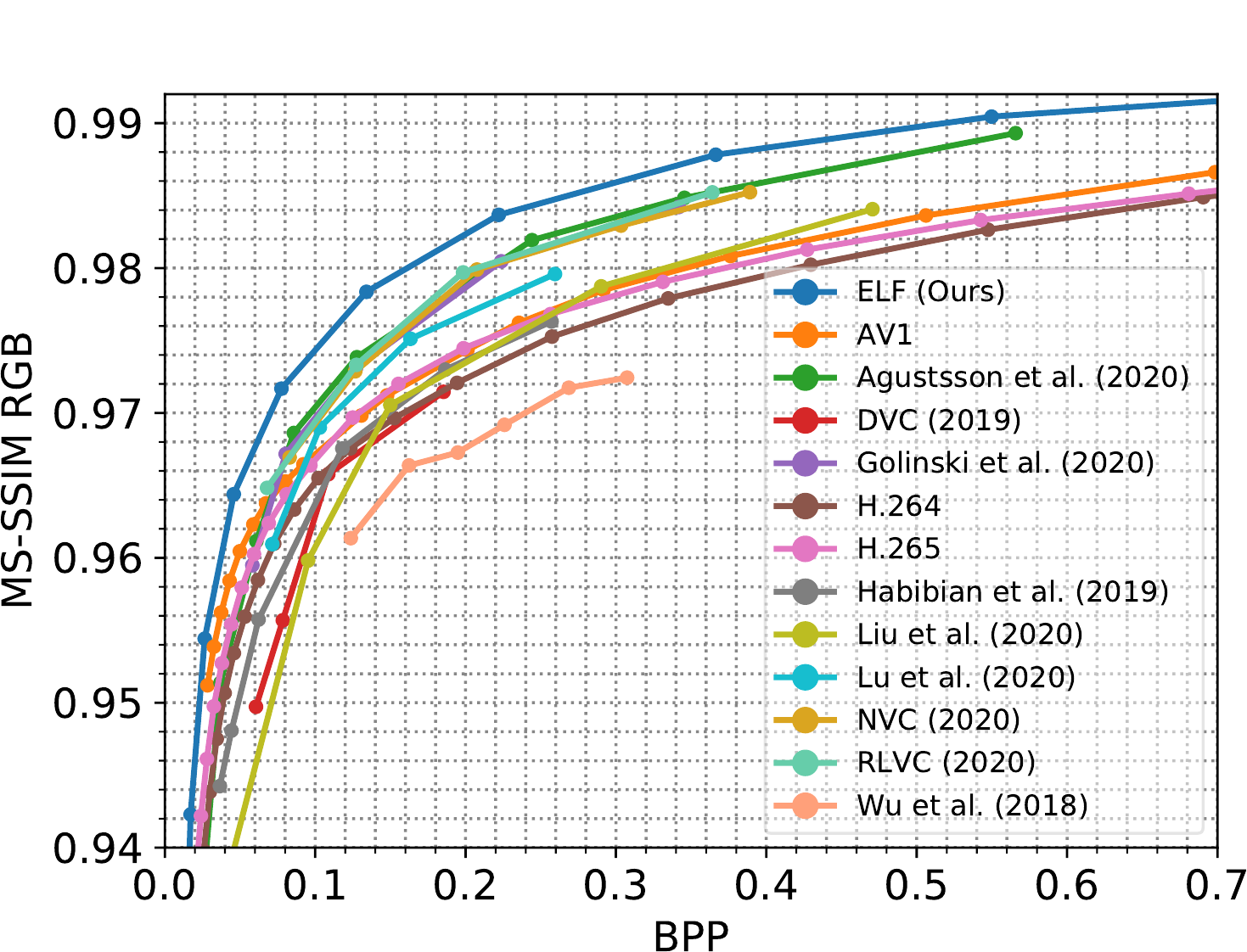}\hspace{-0.02in}
    \includegraphics[height=1.72in]{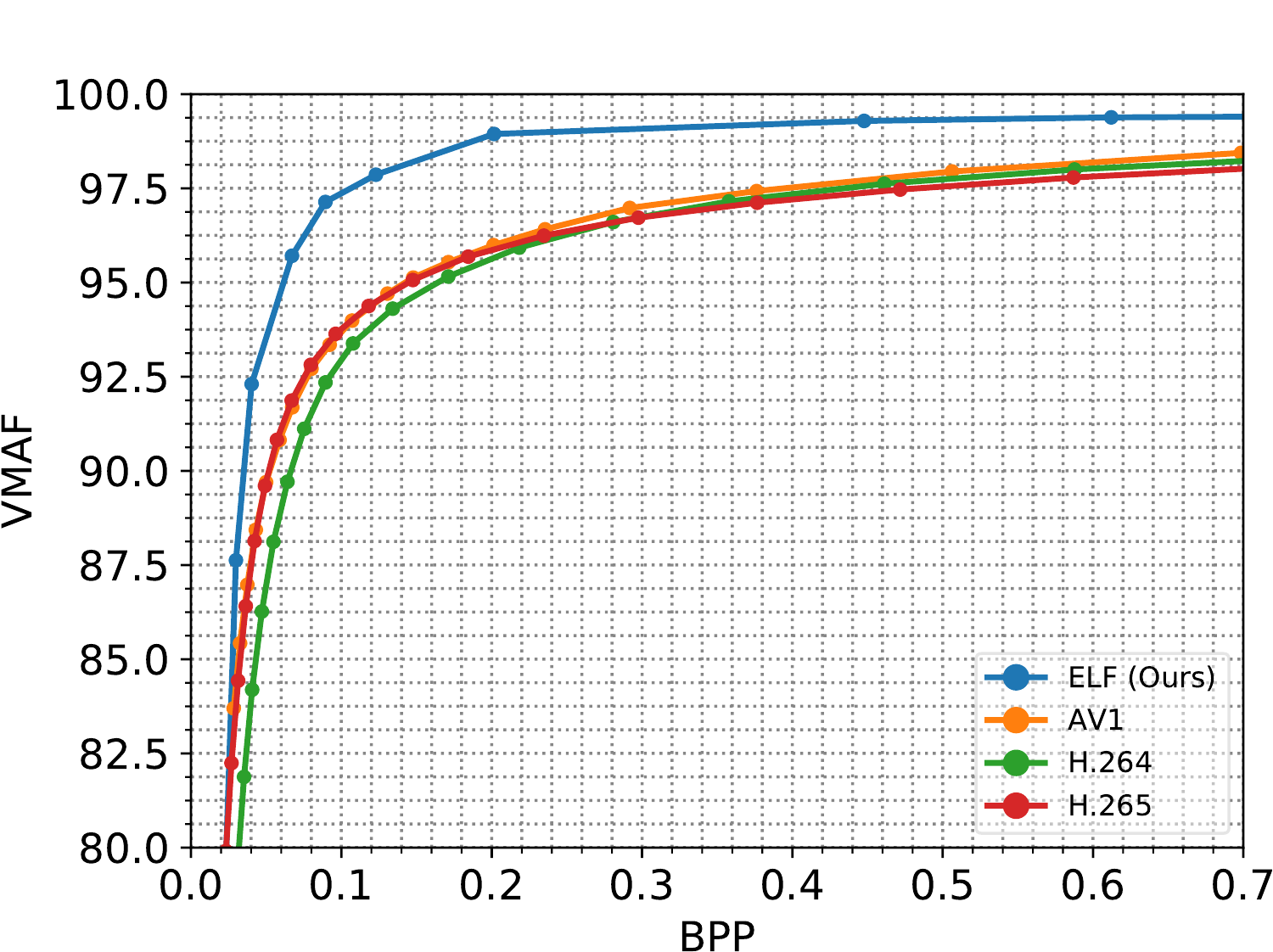}

    \rotatebox[origin=t]{90}{\bf MCL-JCV Dataset\hspace{-1.5in}}\hspace{0.02in}
    \includegraphics[height=1.72in]{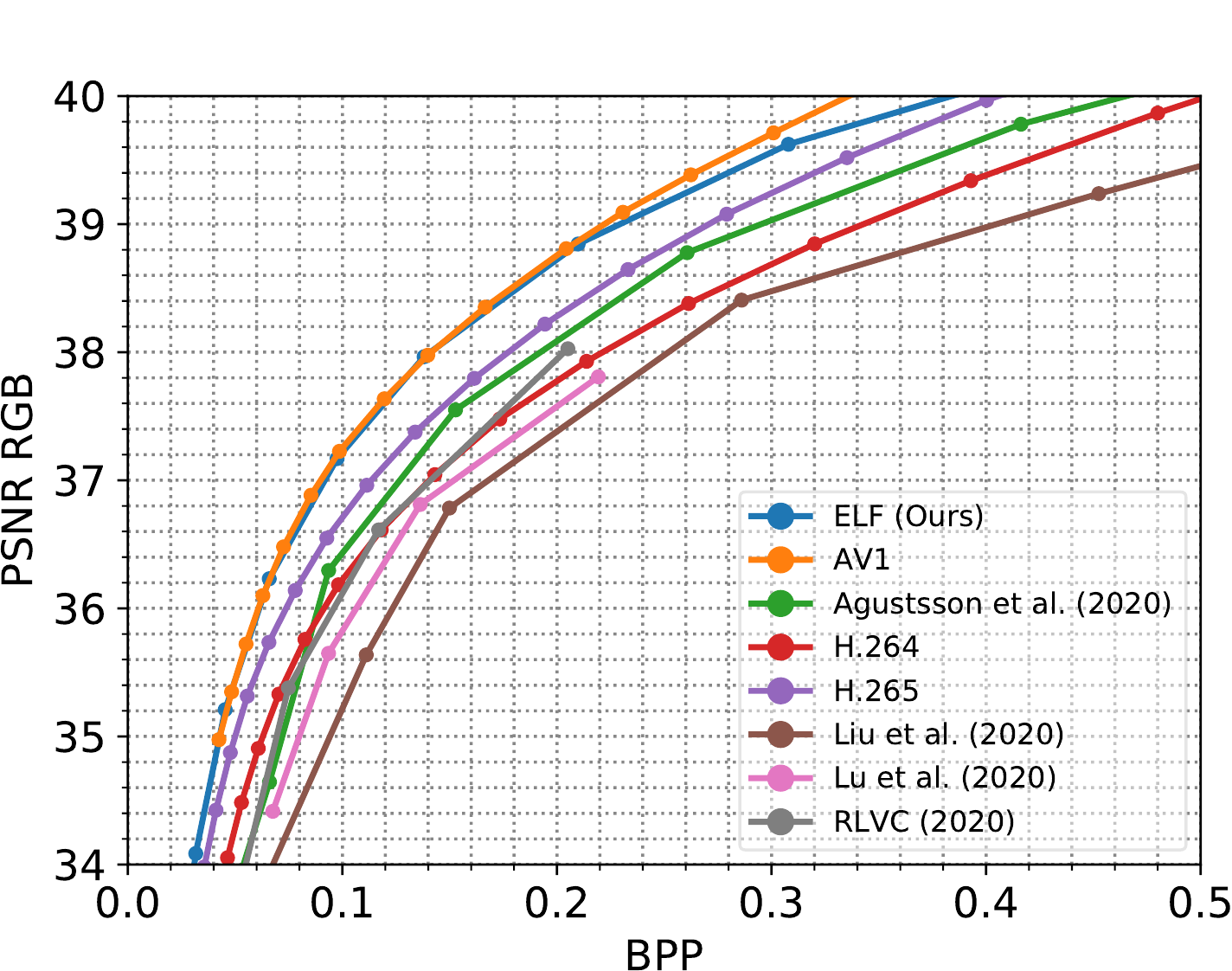}\hspace{-0.02in}
    \includegraphics[height=1.72in]{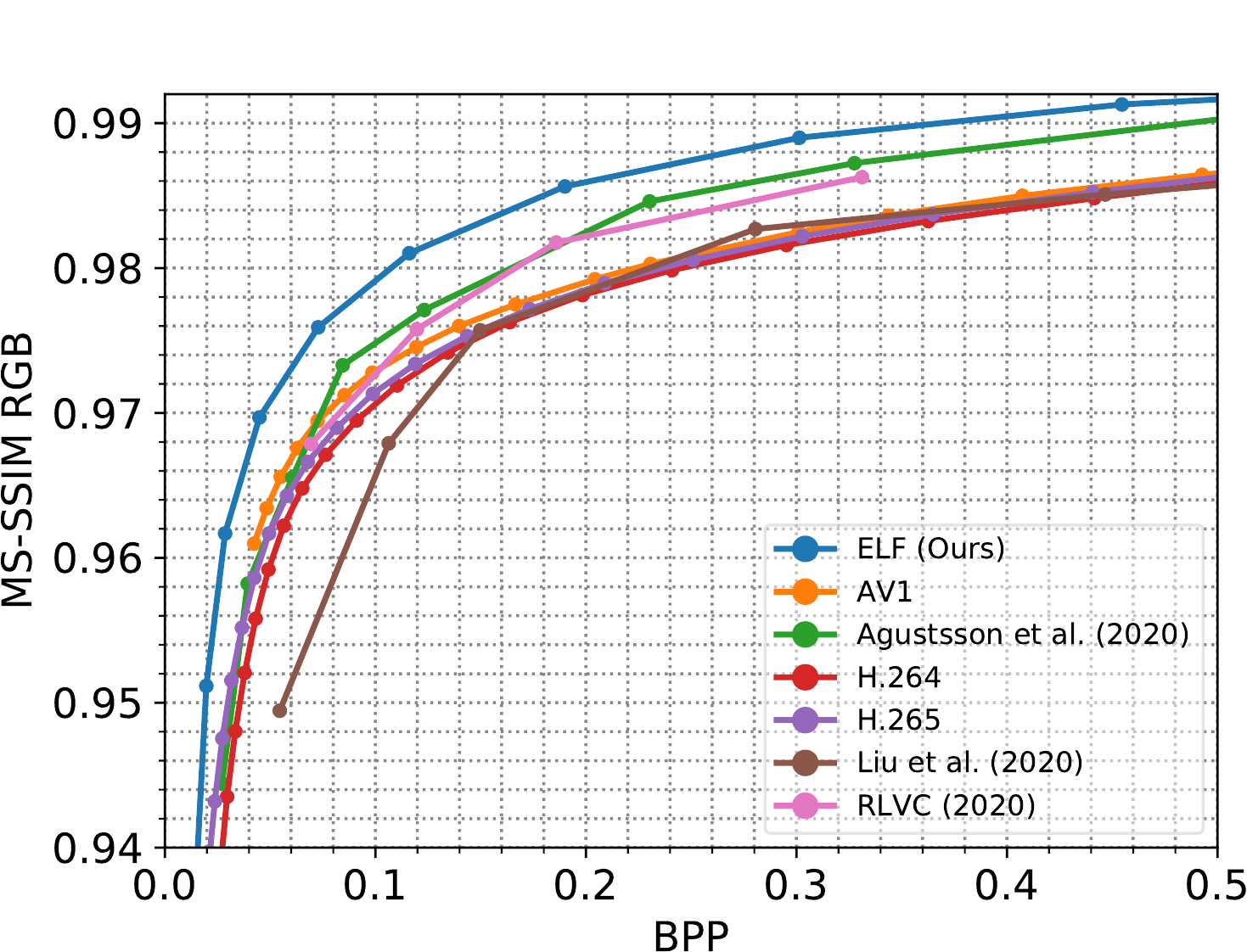}\hspace{-0.02in}
    \includegraphics[height=1.72in]{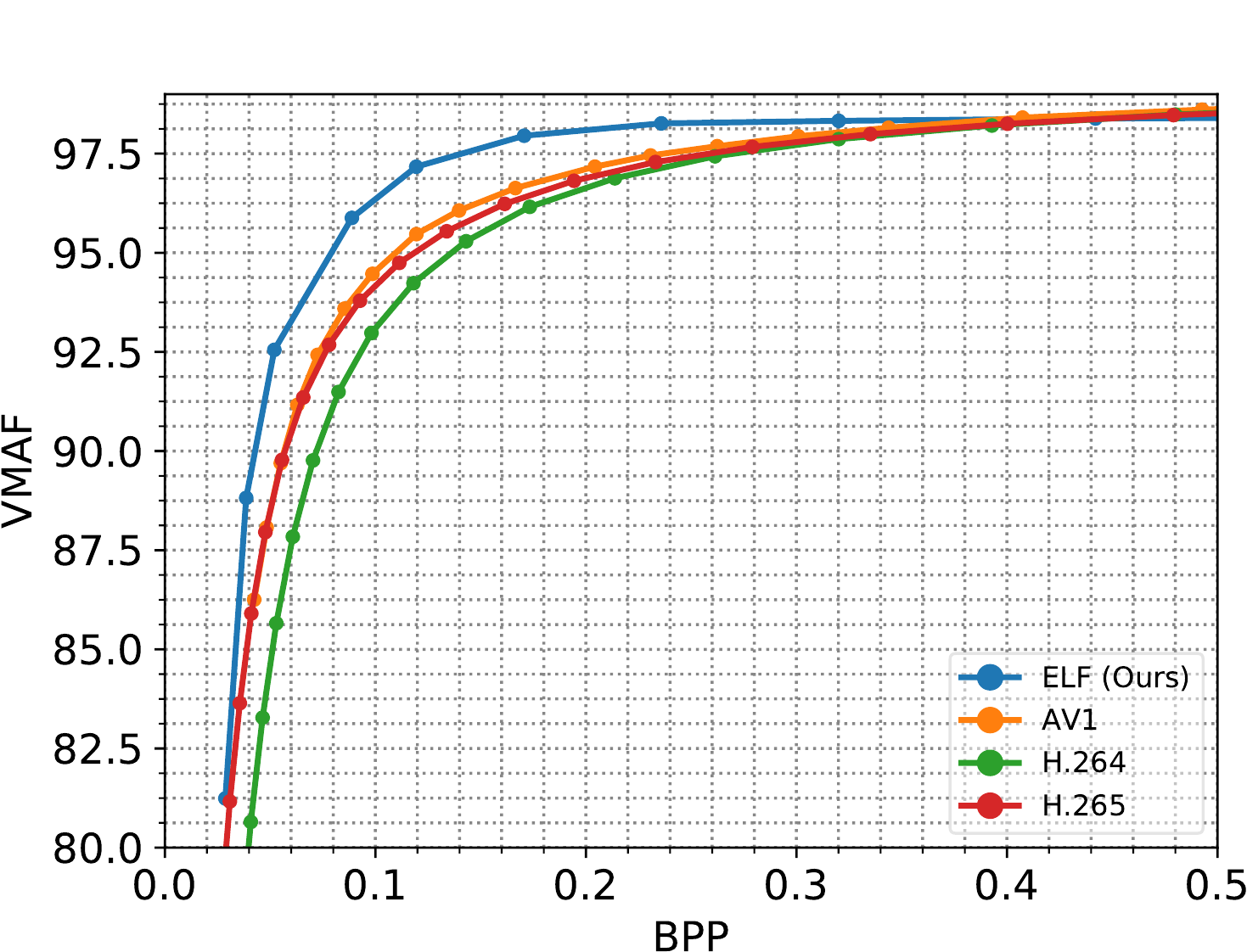}
    
    \caption{Rate-distortion curves of traditional codecs and state-of-the-art ML codecs \cite{agustsson2020scale,lu2019dvc,golinski2020feedback,habibian2019video,liu2020fast,lu2020content,liu2020nvc,yang2020rlvc,wu2018vcii} on the UVG and MCL-JCV video datasets. 
    }
    \label{fig:rd_curves}
    \vspace{-0.2in}
\end{figure*}

Instead, after experimentation (see Appendix) we converge on a mix between residual and dense blocks where several convolutions, each with $C$ channels, are ran with additive accumulation, at which point the intermediate activations are concatenated only a single time and the dimensionality is reduced back down. We refer to such a block as a DM-$C$ block.

In addition to the backbone structure, we also explored various forms of attention popularized for image compression --- but did not find these to enhance coding efficiency.

For increased efficiency, all blocks are exclusively ran in scales at least 4x downsampled relative to pixel space. Detailed descriptions of the end-to-end architectures are found in Appendix \ref{sec:backbone_details}.

\paragraph{Optimizing for a light decoder} 
There exists an inherent asymmetry between encoding and decoding, since in many use-cases the decoder must run in computationally-constrained environments (such as phones). In order to optimize the decoding speed given a fixed encoding speed, we shift computation from the decoder branches (inputs $\shortrightarrow$ codelayer) to the encoder branches (codelayer $\shortrightarrow$ outputs). In our ablations, we find that an asymmetric encoder-decoder pair has better R-D performance in addition to being faster during decoding as compared to a symmetric encoder-decoder pair (Table \ref{tab:ablation}). For both flow and residue encoders we use three DM-256 blocks sequentially in scales 4x-8x-16x. For the flow and residue decoders we use three DM-64 or DM-128 respectively in scale sequence 16x-8x-4x. Complete layer specification is found in Appendix \ref{sec:backbone_details}.

\subsection{In-loop flow prediction}\label{sec:predictor}
We propose a novel \emph{in-loop flow predictor} which predicts the current flow from previously transmitted frames and flows. It does not transmit any additional bits itself, and is ran prior to flow/residual encoding or decoding. It allows for BD-rate savings of 13\% (Table \ref{tab:ablation}).

Intuitively, designing video codecs revolves around exploiting redundancy across frames. Clearly, such similarities are captured well using optical flow; however, we observe that there also exists further redundancy among the optical flow fields \emph{themselves}. Since consecutive flow fields are similar to one another due to linearity for motion, the current flow field can be predicted reasonably well from information already available on the decoder side --- without transmitting any additional information. This predicted flow can then be refined using the flow autoencoder.

The predictor structure can be found in Figure \ref{fig:overall_architecture}. It takes in as inputs the previous flow $\rmbhf_{t-1}$ and two previous frame reconstructions $\rmbhx_{t-2}$ and $\rmbhx_{t-1}$, and produces a base flow prediction $\rmbbf_t$. The flow autoencoder is re-tasked with producing a sigmoid map $\rmbm_t$ masking the predicted flow elementwise, and sparse flow delta $\bDelta\rmbhf_t$ added to the masked predicted flow. The final flow is then $\rmbhf_t = \rmbm_t\odot\rmbbf_t + \bDelta\rmbhf_t$. The original baseline model in Section \ref{sec:baseline_model}, then, is a special case with a predicted flow of zero and no mask. The predictor computation is shared by both the encoder and decoder.

In Figure \ref{fig:intermediate_tensors}, it is seen that the zero-bit predicted flow (top left) is already similar to the final flow (top right). Hence, the flow block only needs to output sparse flow touchups, thereby spending less bits. The neural network backbone used within the predictor is based upon the DM block described in Section \ref{sec:backbone} but it additionally computes features at two scales for each block (see Appendix \ref{sec:backbone_details} for details). 

A similar idea was proposed by \cite{lin2020m}. We note that this approach does not include learned mask $\rmbm_t$ to allow turning the predictor off in areas that are difficult to predict. Also, in this approach \emph{only} the predictor has a runtime of 17 FPS for resolution $320\times 256$. In contrast, our entire decoder, predictor included, runs in 139 FPS for resolution $352\times 288$ (which has 20\% more pixels).

\begin{table}[t]
\scriptsize
\begin{tabular}{@{}lrrrr@{}}
\toprule
\multicolumn{1}{c}{\multirow{2}{*}{Codec}} & \multicolumn{2}{c}{{\ul\qquad\ \ \  PSNR\qquad\ \ \ }} & \multicolumn{2}{c}{{\ul\qquad MS-SSIM\qquad}} \\
\multicolumn{1}{c}{} & \multicolumn{1}{c}{UVG} & \multicolumn{1}{c}{MCL-JCV} & \multicolumn{1}{c}{UVG} & \multicolumn{1}{c}{MCL-JCV} \\ \midrule
H.264 & -44.3\% & -33.7\% & -52.9\% & -51.2\% \\
H.265 & -26.1\% & -17.0\% & -46.0\% & -46.1\% \\
AV1 & -14.8\% & -1.0\% & -50.0\% & -50.9\% \\ \midrule
Agustsson et al. (2020) \cite{agustsson2020scale} & -35.3\% & -29.4\% & -32.6\% & -34.8\% \\
RLVC (2020) \cite{yang2020rlvc} & -35.0\% & -35.3\% & -28.6\% & -38.4\% \\
NVC (2020)\cite{liu2020nvc} & -37.9\% & - & -31.1\% & - \\
Lu et al. (2020) \cite{lu2020content} & - & -42.3\% & -40.4\% & - \\
Liu et al. (2020) \cite{liu2020fast} & -54.2\% & -46.8\% & -60.0\% & -60.7\% \\
DVC (2019) \cite{lu2019dvc} & -47.6\% & - & -59.9\% & - \\
Habibian et al. (2019) \cite{habibian2019video} & -57.3\% & - & -53.6\% & - \\
Wu et al. (2018) \cite{wu2018vcii} & -62.9\% & - & -69.5\% & - \\ \bottomrule
\end{tabular}
\caption{BD-rate savings of ELF relative to common video standards, and state-of-the-art ML codecs (full R-D curves available in Figure \ref{fig:rd_curves}). Numbers are reported on the PSNR and MS-SSIM metrics for the UVG and MCL-JCV datasets.}
\label{tab:bd_rates}
\vspace{-0.2in}
\end{table}

\section{Results}\label{sec:results}
\subsection{Experimental setup}\label{sec:experimental_setup}
We train our models on the Vimeo-90k dataset \cite{xue2019video} and on MSE using the Adam optimizer \cite{kingma2014adam} for a total of $800,000$ iterations with a batch size of 8 and GOP of IPPP. During inference we unroll over P-frames to complete the 16-frame GOP. We train in RGB for PSNR and MS-SSIM, and in YUV for VMAF. We start with a momentum of $0.9$, and learning rate of $7\times 10^{-5}$ which we lower by 5x at 80\% and 95\% into the training. We train for another $40,000$ iterations on a crop size of $320\times 320$ on a Vimeo90k-like dataset we generated with a larger crop size (Appendix \ref{sec:dataset_details}). For models reporting on MS-SSIM or VMAF, we fine-tune for $80,000$ iterations on the respective metrics. 

We train a total of 2 models which together cover the entire bitrate range. During training, each model aims to optimize $L=8$ different points on the R-D curve where the regularization weights $\lambda_{\textrm{reg}}$ are chosen linearly in logspace in the range $[10^{-1.7}, 10^{-3.7}]$ for the lower BPP model and $[10^{-3}, 10^{-5}]$ for the higher. The level for the initial I-frame, $l_0$, is chosen randomly. The following levels are then chosen as $l_t = l_{t-1} + v_t$ rounded to the nearest integer and clipped to the range $[0, L-1]$, where $v_t \sim \mathcal{N}(\mu=0, \sigma=0.5)$ is sampled from a normal distribution. 

To generate all R-D curves in this paper (apart from the maximum bitrate and minimum quality rate controllers in Figure \ref{fig:rd_curves}), we sweep over levels and keep the level constant across all frames of each video.

The model is trained in FP32. The graph is converted to FP16, apart from the subnet decoder where the output probabilities are sensitive to minor perturbations. This does not lead to any reduction in R-D performance. At inference time, we run all models using TensorRT. Entropy encoding and decoding are implemented on CPU and are parallelized over channels and space (see Appendix \ref{sec:codelayer_details}). The entropy coding portion takes about $10\%$ of the total runtime.

\stepcounter{footnote}
\footnotetext{\cite{yang2020rlvc} only reports runtimes for 240p; to enable comparison on CIF the numbers were scaled proportionally to the number of pixels.\label{foot:rlvc}}

\stepcounter{footnote}
\footnotetext{For \cite{liu2020fast}, we only count their fast C++ implementation; the authors further report that the Python interface leads to an overhead of 1,190ms for encoding and 650ms for decoding, but we ignore these counts.\label{foot:liu_aec}}

\begin{table}[t]
\centering
\fontsize{6.3}{7.5}\selectfont
\begin{minipage}[t]{0.4\columnwidth}
\begin{tabular}[t]{@{}lrr@{}}
\toprule
\multicolumn{1}{c}{\multirow{2}{*}{Method}} & \multicolumn{2}{c}{{\ul\qquad\ \ \ FPS\qquad\ \ \ }} \\
\multicolumn{1}{c}{} & \multicolumn{1}{c}{Encode} & \multicolumn{1}{c}{Decode} \\ \midrule
\multicolumn{3}{c}{CIF 352x288} \\ \midrule
RLVC\footref{foot:rlvc} \cite{yang2020rlvc} & 12 & 25 \\
Rippel et al. \cite{rippel2018learned} & 6 & 30 \\
DVC \cite{lu2019dvc} & 25 & 41 \\
\bf ELF (Ours) & \bf  71 & \bf 139 \\ \midrule
\multicolumn{3}{c}{VGA 640x480} \\ \midrule
Rippel et al.  \cite{rippel2018learned} & 2 & 10 \\
\bf ELF (Ours) & \bf 47 & \bf 91 \\ \bottomrule
\end{tabular}
\end{minipage}
\hfill
\begin{minipage}[t]{0.5\columnwidth}
\begin{tabular}[t]{@{}lrr@{}}
\toprule
\multicolumn{1}{c}{\multirow{2}{*}{Method}} & \multicolumn{2}{c}{{\ul\qquad\ \ \ FPS\qquad\ \ \ }} \\
\multicolumn{1}{c}{} & \multicolumn{1}{c}{Encode} & \multicolumn{1}{c}{Decode} \\ \midrule
\multicolumn{3}{c}{HD 720 1280x720} \\ \midrule
Rippel et al. \cite{rippel2018learned} & 0.5 & 3 \\
\bf ELF (Ours) & \bf {19\ \ \ } & \bf 35 \\ \midrule
\multicolumn{3}{c}{HD 1080 1920x1080} \\ \midrule
Habibian et al. \cite{habibian2019video} & 1.5 & $10^{-3.7}$\\
Wu et al. \cite{wu2018vcii} & 2.4 & $10^{-3}$ \\
Rippel et al. \cite{rippel2018learned} & 0.2 & 1.0 \\
DVC \cite{lu2019dvc} & 1.5 & 1.8 \\
Liu et al.\footref{foot:liu_aec} \cite{liu2020fast} & 2.0 & 3.0 \\
\bf ELF (Ours) & \bf {10\ \ \ } & \bf {18\ \ \ } \\ \bottomrule
\end{tabular}
\end{minipage}

\caption{Comparison of runtimes for different resolutions of all ML codecs which report timings, for BPP 0.2. For HD 1080, ELF runs at least 5x faster than other ML codecs.}
\label{tab:runtimes}
\vspace{-0.2in}
\end{table}

\begin{table*}[]
\renewcommand{\arraystretch}{1.2}
\center
\small
\begin{tabular}{clrllll}
\hline
\multirow{2}{*}{Property} & \multicolumn{1}{c}{\multirow{2}{*}{Option}} & \multirow{2}{*}{\begin{tabular}[c]{@{}l@{}}BD-rate \\ Increase\end{tabular}} & \multicolumn{2}{c}{{\ul FPS for HD 720}} & \multicolumn{2}{c}{{\ul\qquad\ \ \# Param\qquad\ \ }} \\
 & \multicolumn{1}{c}{} &  & \multicolumn{1}{c}{Enc.} & \multicolumn{1}{c}{Dec.} & \multicolumn{1}{c}{Enc.} & \multicolumn{1}{c}{Dec.} \\ \hline
 
\multirow{2}{*}{\begin{tabular}[c]{@{}c@{}}Flow \\ predictor\end{tabular}} & {\ul Yes} & 0\% & 19 & 35 & 38M & 11M  \\
& No & \color{BrickRed} 13\% & \color{ForestGreen} 21 (+10\%) & \color{ForestGreen} 37 (+6\%) & \color{ForestGreen} 37M (-3\%) & \color{ForestGreen} 10M (-9\%)  \\ \hline
 
\multirow{2}{*}{\begin{tabular}[c]{@{}c@{}}Bitrate\\ coverage\end{tabular}} & {\ul Range} & 0\% & 19 & 35 & 38M & 11M  \\
& Point & \color{BrickRed} 1\% & 19 & 35 & \color{BrickRed} 220M (+500\%) & \color{BrickRed} 70M (+500\%) \\ \hline
 
\multirow{5}{*}{Backbone} & {\ul DM} & 0\% & 19 & 35 & 38M & 11M  \\ 
& { Common 128} & \color{BrickRed} 76\% &  \color{BrickRed} 18 (-5\%) & \color{BrickRed} 24 (-31\%)  & \color{ForestGreen} 24M (-37\%) & \color{BrickRed} 12M (+9\%) \\
 & { Common 192} & \color{BrickRed} 43\% & \color{BrickRed} 14 (-26\%) & \color{BrickRed} 20 (-43\%)  & \color{ForestGreen} 28M (-18\%) & \color{BrickRed} 14M (+27\%) \\
 & { DM-Symmetric 128} & \color{BrickRed} 31\% & \color{ForestGreen} 20 (+5\%) & \color{BrickRed} 28 (-20\%) & \color{ForestGreen} 25M (-34\%) & \color{BrickRed} 13M (+18\%) \\
 & { DM-Symmetric 192} & \color{BrickRed} 18\% & \color{BrickRed} 15 (-21\%) & \color{BrickRed} 22 (-37\%) & \color{ForestGreen} 35M (-8\%) & \color{BrickRed} 17M (+55\%) \\ \hline
 
\multirow{2}{*}{\begin{tabular}[c]{@{}c@{}}Loss \\ modulation\end{tabular}} & {\ul Yes} & 0\% & 19 & 35 & 38M & 11M  \\ 
& No & \color{BrickRed} 10\% & 19 & 35 & 38M & 11M  \\ \hline
 
 
\multirow{2}{*}{State} & {\ul Yes} & 0\% & 19 & 35 & 38M & 11M \\ 
& No & \color{BrickRed} 37\% & 19 & \color{ForestGreen} 36 (+3\%) & \color{ForestGreen} 37M (-3\%)  & 11M  \\ \hline
\end{tabular}
\caption{Ablation studies of different modeling choices. The BD-rate values were computed under PSNR on UVG.}
\label{tab:ablation}
\vspace{-0.1in}
\end{table*}

\subsection{Benchmarking procedure}
\paragraph{Baseline codecs} 
We benchmark against modern commercial codecs H.264/AVC, H.265/HEVC, and AV1, as well as the most competitive ML codecs to our knowledge. We use FFmpeg to encode H.264 and H.265\footref{foot:google_curves}. We emphasize that while some existing works restrict the baselines to the \texttt{veryfast} preset, we use the default preset, which is a much more competitive and realistic baseline. We do not constrain the codecs in any way apart from disabling B-frames. We use the SVT-AV1 encoder for AV1. See Appendix \ref{sec:standards_commands} for the exact commands used. For the ML-based codecs, we compare against all recent approaches including \cite{wu2018vcii,lu2019dvc,habibian2019video,golinski2020feedback,lu2020content,agustsson2020scale,liu2020nvc,liu2020fast,yang2020rlvc}\footref{foot:habib}. 

\stepcounter{footnote}
\footnotetext{\cite{habibian2019video}'s PSNR results on the UVG dataset were taken from \cite{liu2020fast}.\label{foot:habib}}

\stepcounter{footnote}
\footnotetext{We use the latest FFmpeg to benchmark the standards. Our H.264 and H.265 curves are slightly better than the ones in \cite{agustsson2020scale}, but we validated we do match them with an older FFmpeg version.\label{foot:google_curves}}

\paragraph{Metrics} 
We evaluate all reconstructions on popular video metrics PSNR, MS-SSIM \cite{wang2003multiscale}, and VMAF\footref{foot:vmaf}. We run VMAF using its hook into FFmpeg. In order to compare against existing ML-based approaches, PSNR and MS-SSIM are evaluated in the RGB colorspace. This is not ideal by any means and YUV 4:2:0 is preferable for perceptual quality optimization.

We found that it is easy for ML codecs to perform well on MS-SSIM and VMAF by the numbers, but that perceptual quality is not commensurate with these gains. We took extra measures to avoid overfitting on VMAF by training it jointly with PSNR.

\stepcounter{footnote}
\footnotetext{See repository at \url{https://github.com/Netflix/vmaf}.\label{foot:vmaf}}

\paragraph{Test sets} 
We benchmark all codecs on popular video datasets UVG \cite{mercat2020uvg} and MCL-JCV \cite{wang2016mcl}. These datasets are commonly used for video codec evaluation, and contain respectively 7 and 30 diverse HD 1080 videos with totals of 3,900 and 4,115 frames.

\subsection{Performance}

\paragraph{Coding efficiency}
Figure \ref{fig:rd_curves} presents the rate-distortion curves for all approaches, and Table \ref{tab:bd_rates} provides BD-rate \cite{Bjontegaard2001CalculationOA} summaries of these curves. It can be seen that ELF compares favorably against all standards and ML codecs, under all metrics --- with the exception of BPP (bits per pixel) 0.25 and higher for the MCL-JCV dataset against AV1. In analyzing as to why, we observed that our model, similar to \cite{agustsson2020scale}, performs very poorly on the four non-photorealistic/cartoon videos within MCL-JCV. Excluding these videos we outperform AV1 across the entire range (Appendix \ref{sec:results_by_video}).

\paragraph{Computational efficiency} 
We benchmark the runtime of our codec across different resolutions on an NVIDIA Titan V GPU, and include all time spent on network execution, entropy encoding/decoding, and so on. We only exclude CPU$\leftrightarrow$GPU memory transfer overhead: this is not a fundamental limitation but rather an artifact of TensorFlow's inability to pin data to GPU memory across model iterations. The benchmarks and comparisons against all other approaches which report timings\footref{foot:liu_timings} can be found in Table \ref{tab:runtimes}. For example, for HD 1080, ELF runs at 98ms/frame for encoding and 55ms/frame for decoding; this is 5x faster than the second-fastest ML codec \cite{liu2020fast}, while reducing the BD-rate by 55\% on average relative to it (Table \ref{tab:bd_rates}).

\stepcounter{footnote}
\footnotetext{Many of the timings were taken from \cite{liu2020fast}, who were able to gather these from the original authors of the respective papers. We were further able to contact \cite{rippel2018learned} who generously provided detailed benchmarks. \cite{lu2019dvc,yang2020rlvc,liu2020fast} use an NVIDIA 1080 Ti GPU, and \cite{rippel2018learned} uses an NVIDIA Titan V. \label{foot:liu_timings}}

\subsection{Ablation studies}
\label{sec:ablation_studies}
We study the individual contributions of the proposed ideas. For the ablation environment we follow exactly same training and inference procedures described in \ref{sec:experimental_setup}, apart from training each model for 250,000 iterations and only training the lower bitrate range model. The results can be found in Table \ref{tab:ablation}. 

Using 2 flexible-rate models instead of 12 single-rate models reduces the number of required parameters by 6x without harming compression performance at all.

Swapping to the common backbone often used in the image/video coding literature \cite[\dots]{balle2018variational,agustsson2018generative}, with base channels 128 or 192 ("Common" in the ablation table) worsens BD-rate by at least 43\% and slows the decoder down by 31\% (we do our best to tune the model for this backbone). 

Swapping to a symmetric DM-based backbone where the encoder and decoder have the same number of channels ("Symmetric" in the ablation table) increases the BD-rate by at least 18\% and slows down the decoder by at least 20\%.

Removing the predictor results in a BD-rate increase of 13\%, without changing runtimes dramatically. Removing the dynamic loss modulation worsens BD-rate by 10\%. 

\section{Conclusion}
Compared to the video coding standards, ML-based video compression is still in its infancy. While ML video codecs have achieved impressive R-D performance, some of this success can be attributed to the huge investment in general in neural network research, software, and hardware. There still remains considerable work to be done towards practical deployment of ML codecs.

In this work, we propose ELF-VC, which makes significant improvements in terms of R-D performance, bitrate flexibility and speed. We believe there still exists significant room for improvement across all modeling decisions: architectural choices for the different modules, rate control, and so on.  Moreover, while our approach is close to real-time on a mid-range desktop GPU, it still requires further optimization to achieve real-time performance on edge devices such as phones. Another important direction for future work in ML-based video compression is a visual perception metric for video that is suitable for backpropagation and closely aligns with human visual perception.

{\small
\bibliographystyle{ieee_fullname}
\bibliography{main} 

\begin{thebibliography}{10}\itemsep=-1pt

\bibitem{Guo2020}
{Variable rate image compression with content adaptive optimization}.
\newblock {\em IEEE Computer Society Conference on Computer Vision and Pattern
  Recognition Workshops}, 2020-June:533--537, 2020.

\bibitem{agustsson2020scale}
Eirikur Agustsson, David Minnen, Nick Johnston, Johannes Balle, Sung~Jin Hwang,
  and George Toderici.
\newblock Scale-space flow for end-to-end optimized video compression.
\newblock In {\em Proceedings of the IEEE/CVF Conference on Computer Vision and
  Pattern Recognition}, pages 8503--8512, 2020.

\bibitem{agustsson2018generative}
Eirikur Agustsson, Michael Tschannen, Fabian Mentzer, Radu Timofte, and Luc
  Van~Gool.
\newblock Generative adversarial networks for extreme learned image
  compression.
\newblock {\em arXiv preprint arXiv:1804.02958}, 2018.

\bibitem{balle2016end}
Johannes Ball{\'e}, Valero Laparra, and Eero~P Simoncelli.
\newblock End-to-end optimization of nonlinear transform codes for perceptual
  quality.
\newblock In {\em Picture Coding Symposium (PCS), 2016}, pages 1--5. IEEE,
  2016.

\bibitem{balle2018variational}
Johannes Ball{\'e}, David Minnen, Saurabh Singh, Sung~Jin Hwang, and Nick
  Johnston.
\newblock Variational image compression with a scale hyperprior.
\newblock In {\em International Conference on Learning Representations}, 2018.

\bibitem{Bjontegaard2001CalculationOA}
G. Bjontegaard.
\newblock Calculation of average psnr differences between rd-curves.
\newblock 2001.

\bibitem{chen2017dual}
Yunpeng Chen, Jianan Li, Huaxin Xiao, Xiaojie Jin, Shuicheng Yan, and Jiashi
  Feng.
\newblock Dual path networks.
\newblock In {\em Advances in Neural Information Processing Systems}, pages
  4467--4475, 2017.

\bibitem{chen2018overview}
Yue Chen, Debargha Murherjee, Jingning Han, Adrian Grange, Yaowu Xu, Zoe Liu,
  Sarah Parker, Cheng Chen, Hui Su, Urvang Joshi, et~al.
\newblock An overview of core coding tools in the av1 video codec.
\newblock In {\em 2018 Picture Coding Symposium (PCS)}, pages 41--45. IEEE,
  2018.

\bibitem{Choi2019}
Yoojin Choi, Mostafa El-Khamy, and Jungwon Lee.
\newblock {Variable rate deep image compression with a conditional
  autoencoder}.
\newblock {\em Proceedings of the IEEE International Conference on Computer
  Vision}, 2019-Octob:3146--3154, 2019.

\bibitem{Cui2020}
Ze Cui, Jing Wang, Bo Bai, Tiansheng Guo, and Yihui Feng.
\newblock {G-VAE: A Continuously Variable Rate Deep Image Compression
  Framework}.
\newblock 2020.

\bibitem{djelouah2019neural}
Abdelaziz Djelouah, Joaquim Campos, Simone Schaub-Meyer, and Christopher
  Schroers.
\newblock Neural inter-frame compression for video coding.
\newblock In {\em Proceedings of the IEEE International Conference on Computer
  Vision}, pages 6421--6429, 2019.

\bibitem{svtav1}
Github.
\newblock Scalable video technology for av1 (svt-av1 encoder and decoder).

\bibitem{golinski2020feedback}
Adam Golinski, Reza Pourreza, Yang Yang, Guillaume Sautiere, and Taco~S Cohen.
\newblock Feedback recurrent autoencoder for video compression.
\newblock {\em arXiv preprint arXiv:2004.04342}, 2020.

\bibitem{habibian2019video}
Amirhossein Habibian, Ties~van Rozendaal, Jakub~M Tomczak, and Taco~S Cohen.
\newblock Video compression with rate-distortion autoencoders.
\newblock In {\em Proceedings of the IEEE International Conference on Computer
  Vision}, pages 7033--7042, 2019.

\bibitem{he2016deep}
Kaiming He, Xiangyu Zhang, Shaoqing Ren, and Jian Sun.
\newblock Deep residual learning for image recognition.
\newblock In {\em Proceedings of the IEEE conference on computer vision and
  pattern recognition}, pages 770--778, 2016.

\bibitem{hu2020ECCV}
Zhihao Hu, Zhenghao Chen, Dong Xu, Guo Lu, Wanli Ouyang, and Shuhang Gu.
\newblock Improving deep video compression by resolution-adaptive flow coding.
\newblock In {\em European Conference in Computer Vision (ECCV)}, 2020.

\bibitem{huang2018multiscale}
Gao Huang, Danlu Chen, Tianhong Li, Felix Wu, Laurens van~der Maaten, and
  Kilian Weinberger.
\newblock Multi-scale dense networks for resource efficient image
  classification.
\newblock In {\em International Conference on Learning Representations}, 2018.

\bibitem{huang2017densely}
Gao Huang, Zhuang Liu, Laurens Van Der~Maaten, and Kilian~Q Weinberger.
\newblock Densely connected convolutional networks.

\bibitem{johnston2019computationally}
Nick Johnston, Elad Eban, Ariel Gordon, and Johannes Ball{\'e}.
\newblock Computationally efficient neural image compression.
\newblock {\em arXiv preprint arXiv:1912.08771}, 2019.

\bibitem{kingma2014adam}
Diederik Kingma and Jimmy Ba.
\newblock Adam: A method for stochastic optimization.
\newblock {\em arXiv preprint arXiv:1412.6980}, 2014.

\bibitem{netflix}
LDV.
\newblock 2017 on netflix - a year in bingeing.
\newblock 2017.

\bibitem{ldv2017}
LDV.
\newblock 45 billion cameras by 2022 fuel business opportunities.
\newblock 2017.

\bibitem{lin2020m}
Jianping Lin, Dong Liu, Houqiang Li, and Feng Wu.
\newblock M-lvc: Multiple frames prediction for learned video compression.
\newblock In {\em Proceedings of the IEEE/CVF Conference on Computer Vision and
  Pattern Recognition}, pages 3546--3554, 2020.

\bibitem{liu2019neural}
Haojie Liu, Tong Chen, Ming Lu, Qiu Shen, and Zhan Ma.
\newblock Neural video compression using spatio-temporal priors.
\newblock {\em arXiv preprint arXiv:1902.07383}, 2019.

\bibitem{liu2020nvc}
Haojie Liu, M. Lu, Zhan Ma, Fan Wang, Zhihuang Xie, Xun Cao, and Yao Wang.
\newblock Neural video coding using multiscale motion compensation and
  spatiotemporal context model.
\newblock {\em ArXiv}, abs/2007.04574, 2020.

\bibitem{liu2020unified}
Jiaheng Liu, Guo Lu, Zhihao Hu, and Dong Xu.
\newblock A unified end-to-end framework for efficient deep image compression,
  2020.

\bibitem{liu2020fast}
Jerry Liu, Shenlong Wang, W. Ma, Meet Shah, Rui Hu, Pranaab Dhawan, and R.
  Urtasun.
\newblock Conditional entropy coding for efficient video compression.
\newblock {\em European Conference on Computer Vision (ECCV)}, 2020.

\bibitem{lu2020content}
Guo Lu, Chunlei Cai, Xiaoyun Zhang, Li Chen, Wanli Ouyang, Dong Xu, and Zhiyong
  Gao.
\newblock Content adaptive and error propagation aware deep video compression.
\newblock {\em arXiv preprint arXiv:2003.11282}, 2020.

\bibitem{lu2019dvc}
Guo Lu, Wanli Ouyang, Dong Xu, Xiaoyun Zhang, Chunlei Cai, and Zhiyong Gao.
\newblock Dvc: An end-to-end deep video compression framework.
\newblock In {\em Proceedings of the IEEE Conference on Computer Vision and
  Pattern Recognition}, pages 11006--11015, 2019.

\bibitem{mercat2020uvg}
Alexandre Mercat, Marko Viitanen, and Jarno Vanne.
\newblock Uvg dataset: 50/120fps 4k sequences for video codec analysis and
  development.
\newblock In {\em Proceedings of the 11th ACM Multimedia Systems Conference},
  pages 297--302, 2020.

\bibitem{mukherjee2013latest}
Debargha Mukherjee, Jim Bankoski, Adrian Grange, Jingning Han, John Koleszar,
  Paul Wilkins, Yaowu Xu, and Ronald Bultje.
\newblock The latest open-source video codec vp9-an overview and preliminary
  results.
\newblock In {\em 2013 Picture Coding Symposium (PCS)}, pages 390--393. IEEE,
  2013.

\bibitem{park2019deep}
Woonsung Park and Munchurl Kim.
\newblock Deep predictive video compression with bi-directional prediction.
\newblock {\em arXiv preprint arXiv:1904.02909}, 2019.

\bibitem{rippel17}
Oren Rippel and Lubomir Bourdev.
\newblock Real-time adaptive image compression.
\newblock In Doina Precup and Yee~Whye Teh, editors, {\em Proceedings of the
  34th International Conference on Machine Learning}, volume~70 of {\em
  Proceedings of Machine Learning Research}, pages 2922--2930, International
  Convention Centre, Sydney, Australia, 06--11 Aug 2017. PMLR.

\bibitem{rippel2018learned}
Oren Rippel, Sanjay Nair, Carissa Lew, Steve Branson, Alexander~G Anderson, and
  Lubomir Bourdev.
\newblock Learned video compression.
\newblock {\em arXiv preprint arXiv:1811.06981}, 2018.

\bibitem{sandler2018mobilenetv2}
Mark Sandler, Andrew Howard, Menglong Zhu, Andrey Zhmoginov, and Liang-Chieh
  Chen.
\newblock Mobilenetv2: Inverted residuals and linear bottlenecks.
\newblock In {\em Proceedings of the IEEE conference on computer vision and
  pattern recognition}, pages 4510--4520, 2018.

\bibitem{sullivan2012overview}
Gary~J Sullivan, Jens-Rainer Ohm, Woo-Jin Han, and Thomas Wiegand.
\newblock Overview of the high efficiency video coding (hevc) standard.
\newblock {\em IEEE Transactions on circuits and systems for video technology},
  22(12):1649--1668, 2012.

\bibitem{facebook}
TechCrunch.
\newblock Facebook hits 100m hours of video watched a day.
\newblock 2018.

\bibitem{toderici2015variable}
George Toderici, Sean~M O'Malley, Sung~Jin Hwang, Damien Vincent, David Minnen,
  Shumeet Baluja, Michele Covell, and Rahul Sukthankar.
\newblock Variable rate image compression with recurrent neural networks.
\newblock {\em arXiv preprint arXiv:1511.06085}, 2015.

\bibitem{wang2016mcl}
Haiqiang Wang, Weihao Gan, Sudeng Hu, Joe~Yuchieh Lin, Lina Jin, Longguang
  Song, Ping Wang, Ioannis Katsavounidis, Anne Aaron, and C-C~Jay Kuo.
\newblock Mcl-jcv: a jnd-based h. 264/avc video quality assessment dataset.
\newblock In {\em 2016 IEEE International Conference on Image Processing
  (ICIP)}, pages 1509--1513. IEEE, 2016.

\bibitem{wang2003multiscale}
Zhou Wang, Eero~P Simoncelli, and Alan~C Bovik.
\newblock Multiscale structural similarity for image quality assessment.
\newblock In {\em Signals, Systems and Computers, 2004.}, volume~2, pages
  1398--1402. Ieee, 2003.

\bibitem{wiegand2003overview}
Thomas Wiegand, Gary~J Sullivan, Gisle Bjontegaard, and Ajay Luthra.
\newblock Overview of the h. 264/avc video coding standard.
\newblock {\em IEEE Transactions on circuits and systems for video technology},
  13(7):560--576, 2003.

\bibitem{wu2018vcii}
Chao-Yuan Wu, Nayan Singhal, and Philipp Kr{\"a}henb{\"u}hl.
\newblock Video compression through image interpolation.
\newblock In {\em ECCV}, 2018.

\bibitem{xue2019video}
Tianfan Xue, Baian Chen, Jiajun Wu, Donglai Wei, and William~T Freeman.
\newblock Video enhancement with task-oriented flow.
\newblock {\em International Journal of Computer Vision (IJCV)},
  127(8):1106--1125, 2019.

\bibitem{yang2020rlvc}
Ren Yang, Fabian Mentzer, Luc Van~Gool, and Radu Timofte.
\newblock Learning for video compression with recurrent auto-encoder and
  recurrent probability model.
\newblock {\em arXiv preprint arXiv:2006.13560}, 2020.

\bibitem{youtube}
YouTube.
\newblock Youtube for press.
\newblock 2020.

\end{thebibliography}
}

\clearpage
\renewcommand{\thesubsection}{\Alph{subsection}}

\subsection{Codelayer Details}
\label{sec:codelayer_details}
The general strategy is to use a convolutional neural network to transform some collection of features into a quantized codelayer, and then to use entropy coding to losslessly encode those features. Another neural network is then used to decode the codelayer and produce tensors of interest.  An entropy coder can efficiently compress a tensor if a well-calibrated probability for each possible value of that tensor is provided.  For the I-frame, the input is the image to be compressed. For the P-frame, the input includes other features such as the previous flow and previous reconstruction. The architecture uses side information ($\rmbq_1$) to encode the main codelayer ($\rmbq_0$) more efficiently. 

In equations, we have $\rmbq_0 = \rmbQ(\rmbE_0(\text{inputs}))$, $\rmbq_1=\rmbQ(\rmbE_1(\rmbq_0))$, 
$\mu_0, \sigma_0 = \rmbD_1(\rmbq_1)$, and $\text{outputs} = \rmbD_0(\rmbq_0)$ where $\rmbE_0, \rmbE_1, \rmbD_1, \rmbD_0$ are convolutional neural networks and $\rmbQ(\rmbx) = Q_w\text{round}(\rmbx /(Q_w)) $ is a point-wise quantization function with quantization width $Q_w$. $\rmbq_0$ and $\rmbq_1$ are tensors of the shape $H/s_i, W/s_i, C$ where $H,W$ are the dimensions of the frame, $s_i$ is an integer stride, and $C$ is a number of channels. $\bmu_0, \bsigma_0$ are tensors of the same shape as $\rmbq_0$. They assign a mean and standard deviation for each element of the codelayer. 

In order to use entropy coding to losslessly encode the discrete values, $q\in \{0, \pm Q_w, \pm 2Q_w, \ldots \}$, in the quantized codelayers $\rmbq_0$ and $\rmbq_1$, a probability for each possible discrete value needs to be assigned. A Gaussian with support on the real line can be used to provide probabilities for a discrete distribution by computing the area under the Gaussian within $\pm 1/2$ of the quantization width: 
$p(q|\mu, \sigma, Q_w)= \int_{q-Q_w/2}^{q+Q_w/2} \mathcal{N}(x|\mu, \sigma)\, dx = \Phi(\frac{q - \mu + Q_w/2}{\sigma}) - \Phi(\frac{q - \mu - Q_w/2}{\sigma})$  where $\Phi$ is the CDF of the standard normal distribution.  

The codelength can be computed and included as part of the differentiable loss by computing the sum over all elements in the codelayers $\sum_{i, j}-\log_2 p(q_{i, j}|\mu_i, \sigma_i, Q_w)$ where $j$ sums over the height, width, and channel axes of the codelayer tensor. This interpretation of the probabilities for each discrete quantized value results in the same equations as Balle \textit{et al.}, but is simpler than their interpretation which involves convolving an initial density model with a uniform distribution. 

Since the quantized tensors are of a shape a $H'$ by $W'$ by $C$, an estimate of the codelength as a \textit{function of space} can be computed by summing the log probabilities over the channel axis, using nearest neighbor upsampling to align the codelength maps which have different strides. 

A Gaussian range encoder ($\text{GRE}$) is used to losslessly encode the quantized codelayer $\rmbq_0$ and subnet codelayer $\rmbq_1$. During encoding: $\rmbb_1 = \text{GRE}(\rmbq_1|\bmu=0, \bsigma=\bsigma_1)$ and $\rmbb_0=\text{GRE}(\rmbq_0|\bmu=\bmu_0, \bsigma=\bsigma_0)$. During decoding a Gaussian range decoder ($\text{GRD}$) is used to decode the encoded bits: 
$\rmbq_1 = \text{GRD}(\rmbb_1|\bmu=0, \bsigma=\bsigma_1)$ and $\rmbq_0 = \text{GRD}(\rmbb_0|\bmu=\bmu_0, \bsigma=\bsigma_0)$. $\bsigma_1$ is a parameter of the model and gives a standard deviation for each channel of $\rmbq_1$. 

Since the Gaussian probability model has a diagonal covariance matrix, the elements of the codelayer can be encoded in parallel. Autoregressive probability models, which are common in ML-based compression, are not used because they result in prohibitively slow decoding in their current form.

\subsection{Backbone Architecture}
\label{sec:backbone_details}
\begin{figure*}[t]
    \centering
    \includegraphics[height=\textwidth]{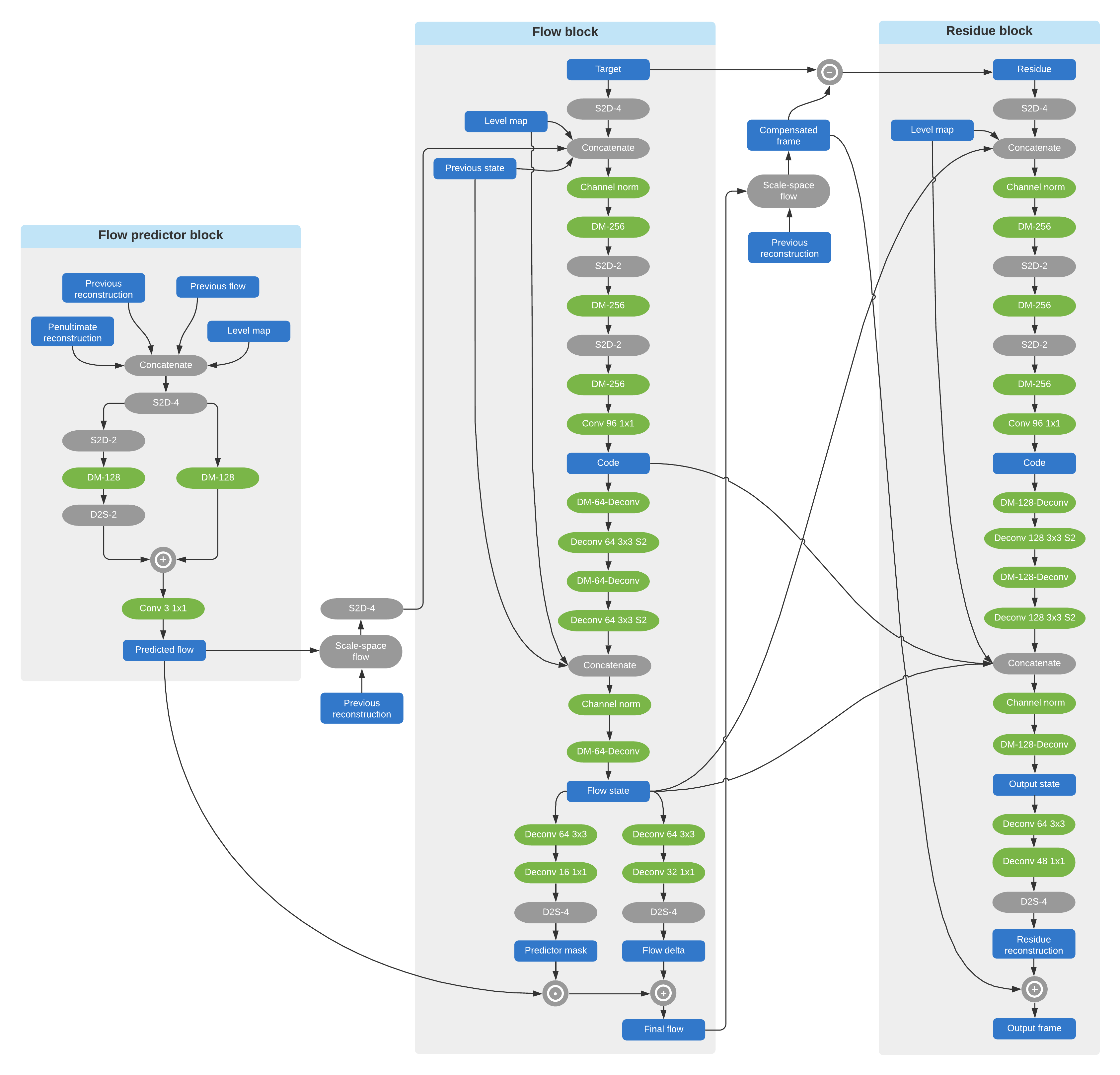}
    \caption{Full layer specification of our model along with backbone configurations and DM block parameters.}
    \label{fig:detailed_architecture}
\end{figure*}

Full layer specification of our model along with backbone configurations and DM block parameters can be found in Figure \ref{fig:detailed_architecture}.

\subsection{Channel Normalization}
In a number of locations of the network, tensors with different average magnitudes are concatenated across the channel axis and fed as input to a subsequent neural network. For instance, the flow is in units of pixels and may be on the order of $10$. In contrast, the residual is normalized to roughly be between $-1$ and $1$. In standard neural networks, normalization is handled by batch normalization. 

However, one issue with batchnorm is that it effectively injects noise into the network, as the output for a particular example depends on the other randomly sampled elements in the batch. Adding noise during training in this way is not ideal for the problem of compression. Instead, we compute channel-wise moving averages of the means and variances of the input to the layer. Then those features are normalized using the computed means and variances. In order to get around the issue of the mean and variance drifting to infinity, we freeze the moving averages after $2000$ training steps. In practice, failing to normalize the inputs in one way or another causes the network to either blow up or to under-perform significantly.

\subsection{Loss Modulator}
\label{sec:extended_loss_modulator}

The loss modulator multiplies the reconstruction loss for a given P-frame and level by a factor $\mu$ in order to give extra weight to modes that are under-performing during training. 

Since the training uses the MSE, the difference in PSNR criterion is converted into a multiplicative factor based on the MSE using the following relationship:  $\text{PSNR}_1 > \text{PSNR}_2 - \delta \Longleftrightarrow \text{MSE}_1 < \text{MSE}_2 \cdot f(\delta)$ where $f(\delta) = 10^{\delta / 10}$. Empirically, we used $f(\delta) = 1.5 \implies \delta = 1.76$. The threshold for the loss modulator was chosen by training the model for 30 epochs for a few different values of the $\delta$ between the PSNR of the I- and P-frames. 
$\mu$ is initialized to $1$ for all frames and levels, but isn't particularly sensitive to the initialization. $\mu$ is clipped to be in the range $[1.0, 10.0]$ to improve training stability. 

This method can also be seen as annealing the regularization weight to allow for larger bitrates for under-performing frames and as the performance improves, the bitrate is more aggressively regularized. We hypothesize that this method is useful because if the regularization is too strong, the model can struggle to train. In equations, the loss can be written as:
\begin{align}
\sum_{l, t} \mu_t^{(l)} \cdot \left[L_{\textrm{rec}, t}^{(l)} + \frac{\lambda_{\textrm{reg}}^{(l)}}{\mu_t^{(l)}} R_t^{(l)}\right]
\end{align}
Thus the effective regularization weight $\frac{\lambda_{\textrm{reg}}^{(l)}}{\mu_t^{(l)}}$ is smaller when $\mu$ is increased. 

In practice, the $\mu_t^{(l)}$ as a function of training iteration starts at $10$, stays there, and then more or less smoothly transitions to $1$.  The lower levels (small $l$) and earlier frames (small $t$) train more efficiently so the crossover time is earlier for those frames and levels. 

\subsection{Level interpolation}
\label{sec:rate_interpolation}

Suppose that one wants to embed $L$ levels in a $L_e$-dimensional space in a way that smoothly interpolates between levels and reduces to a one-hot representation when $L_e=L$.  The embedding of a level $l\in\{0, 1, \ldots L - 1\}$ is computed as follows:
\begin{align}
    s &= \frac{l (L_e - 1)}{L - 1} \\
    u &= \lfloor s\rfloor, \qquad v = u + 1 \\
    d_u &= s - u,\qquad d_v = v - s \\
    \alpha &= \frac{d_v}{d_u + d_v}, \qquad \beta = 1 - \alpha\\
    v_l &= \alpha\cdot\text{onehot}(u|L_e) + \beta \cdot\text{onehot}(v|L_e)
\end{align}

\begin{figure}[b]
    \centering
    \includegraphics[width=\columnwidth]{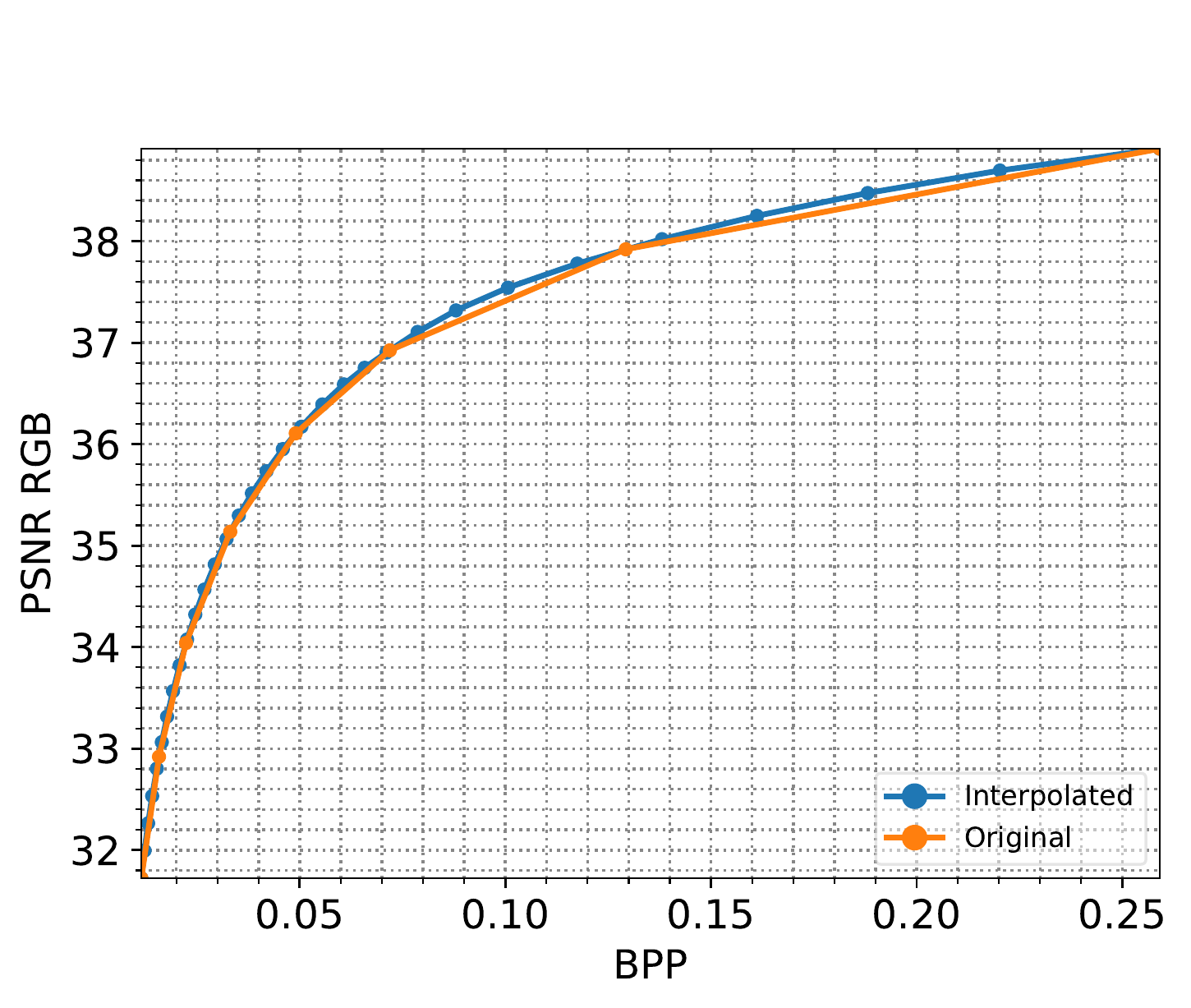}
    \caption{R-D curves of the baseline and level-interpolated model match on the UVG dataset. The baseline model has 8 levels and the interpolated model supports 32 levels.}
    \label{fig:rate_interpolation}
\end{figure}

$\text{onehot}(a|b)$ gives the $b$-dimensional one-hot representation of the integer $a$. If $a<0$ or $a\ge b$, it returns the zero vector. It can be verified that if $L_e=L$, this reduces back to the original one-hot representation. When $L > L_e$, this method smoothly interpolates between one-hot vectors.  This embedding allows us to arbitrarily increase the number of levels without additional training (Figure \ref{fig:rate_interpolation}).  The level-interpolated model is used in the rate control figures in the main text. 

\subsection{Dataset}
\label{sec:dataset_details}
The Vimeo90k dataset \cite{xue2019video} consists of 91,701 7-frame sequences with fixed resolution $448 \times 256$. For finetuning with a larger crop size, we generate a Vimeo90k-like dataset, which consists of 100k 32-frame clips of resolution $352 \times 352$. The clips were generated from 2363 full length original videos which were used to generate the Vimeo90k dataset (\url{http://data.csail.mit.edu/tofu/dataset/original_video_list.txt}). The original videos were pre-processed into distinct segments with a basic threshold-based scene-cut detector. The 32-frame clips were then extracted from the segments at a random downscale factors to ensure a wide range of motion in the dataset.

\subsection{Results by Video Type}
\label{sec:results_by_video}
\begin{figure}[t]
    \centering
    \includegraphics[width=\columnwidth]{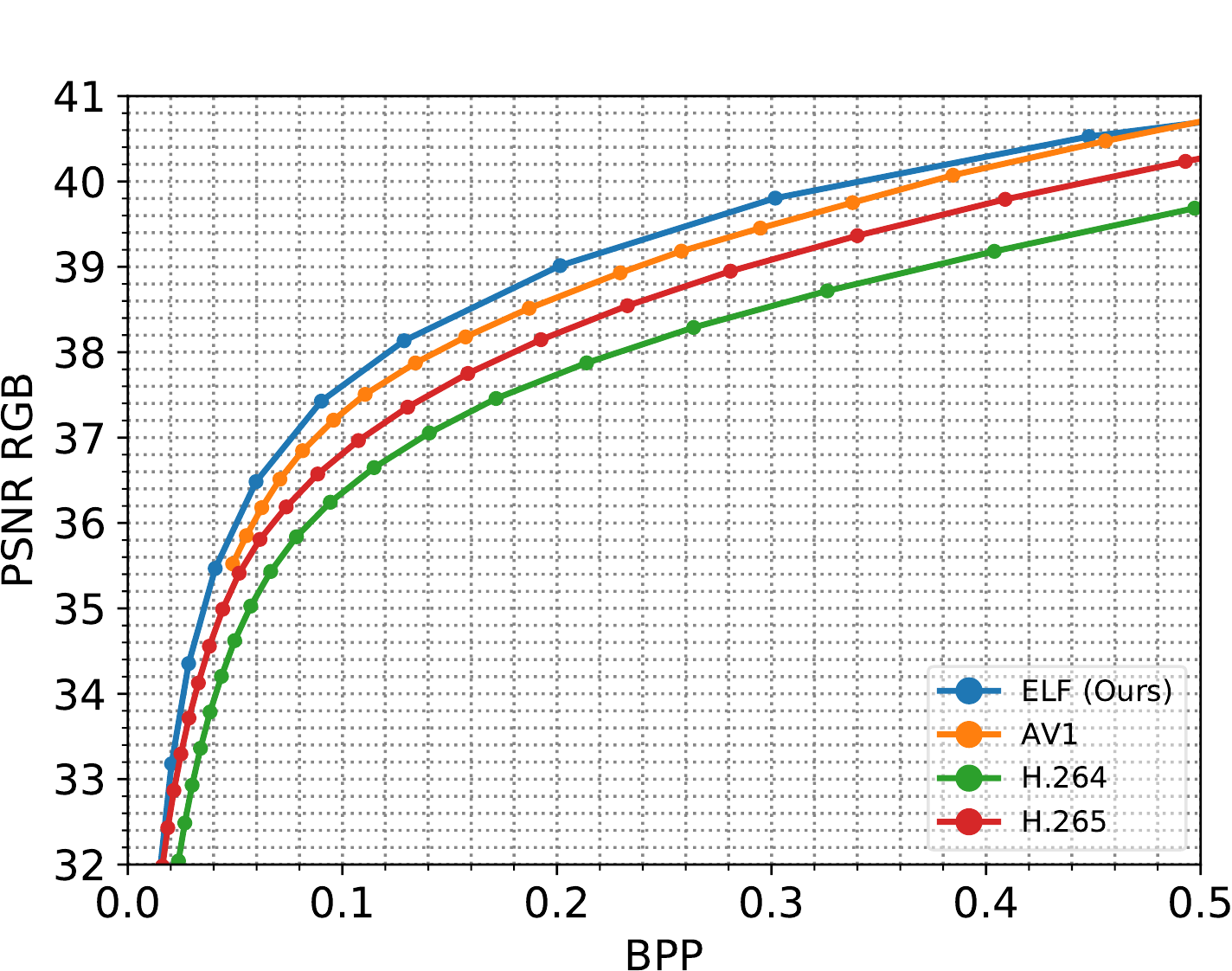}
    \caption{ELF R-D curves on all natural videos in the MCL-JCV dataset.}
    \label{fig:mcl_photorealistic}
\end{figure}

\begin{figure}[t]
    \centering
    \includegraphics[width=\columnwidth]{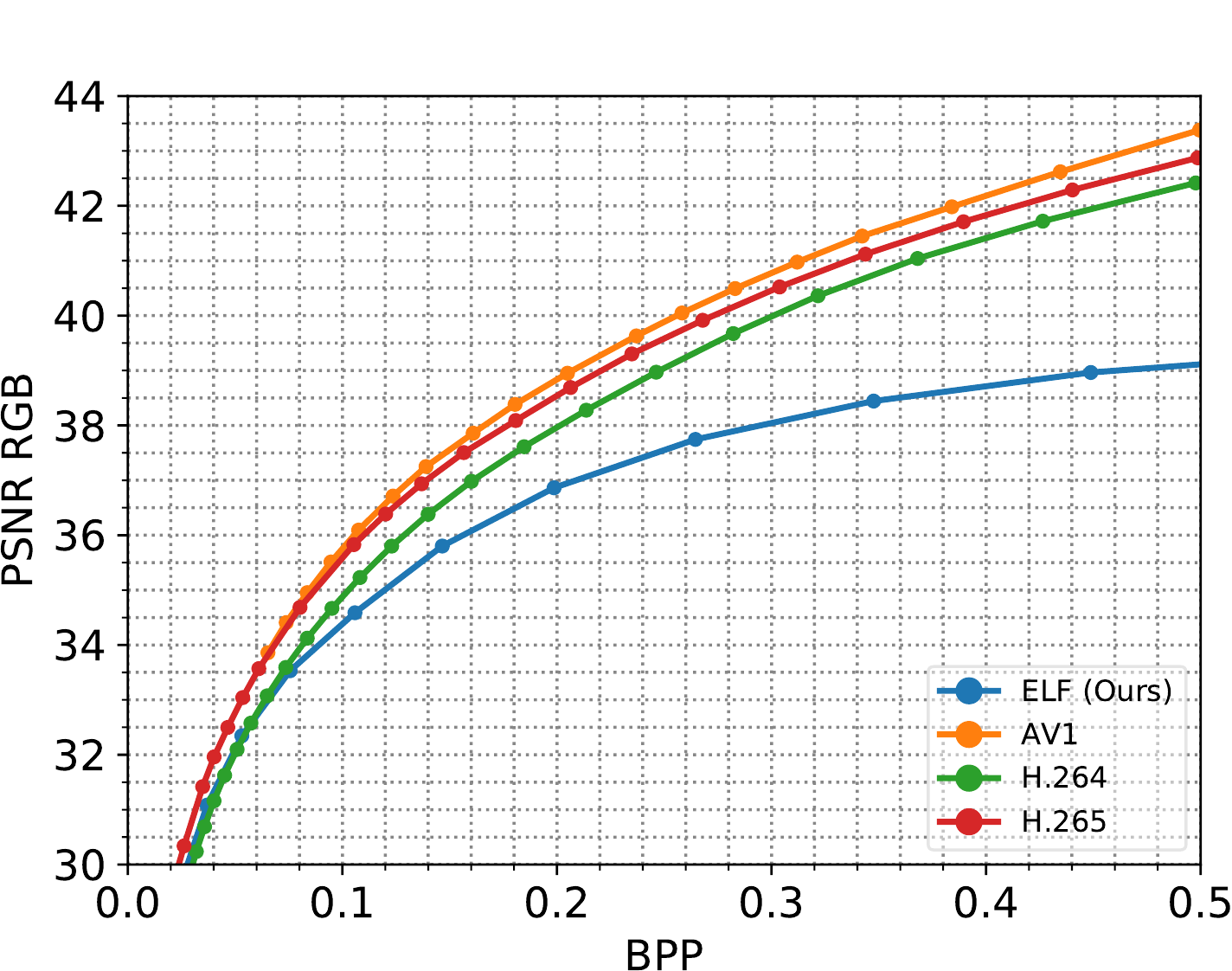}
    \caption{ELF R-D curves on the four non-photorealistic videos in the MCL-JCV dataset.}
    \label{fig:mcl_cartoon}
\end{figure}

As mentioned in the main paper, we found that our codec suffers on non-photorealistic videos. In Figure \ref{fig:mcl_cartoon} we plot the performance of ELF on the four non-photorealistic videos in the MCL-JCV dataset (video IDs 18, 20, 24, 25 within the set), and in Figure \ref{fig:mcl_photorealistic} plot on all other videos.

\subsection{Commands Used for Standards-based Video Compression}
\label{sec:standards_commands}
\paragraph{H.264}
We use the following command to encode all H.264 videos in the paper:
\begin{verbatim}
ffmpeg -i [SRC] \
    -preset medium \
    -codec:v libx264 \
    -crf [RATE] \
    -x264-params bframes=0 [DST]
\end{verbatim}

\paragraph{H.265}
We use the following command to encode all H.265 videos in the paper:
\begin{verbatim}
ffmpeg -i [SRC] \
    -preset medium \ 
    -codec:v libx265 \ 
    -crf [RATE] \ 
    -x265-params bframes=0 [DST]
\end{verbatim}

\paragraph{AV1}
We use the SVT-AV1 Encoder \cite{svtav1} (version v0.8.5-72-gd210088) for comparison. Encoding using only I– and P-frame types is disabled by default in the SVT-AV1 encoder, but can be enabled as mentioned in the github issue \url{https://github.com/AOMediaCodec/SVT-AV1/issues/973}. The encoding command used is: 

\begin{verbatim}
SvtAv1EncApp -i [SRC] \ 
    -b [DST] \ 
    --rc 0 -q [RATE] \ 
    --hierarchical-levels 0 \ 
    --lookahead 0
\end{verbatim}

\subsection{Model Consolidation}
We start with two separate R-D curves, one for each model. Given the selection of regularization weights, the two curves cover different BPP ranges, but still overlap. 

These two R-D curves are consolidated together into a single one. This is simply done by computing the upper convex hull of all R-D points, and keeping the points which are used to construct this hull.

\end{document}